\documentclass[fleqn,usenatbib]{mnras}

\usepackage{newtxtext,newtxmath}
\usepackage[T1]{fontenc}
\usepackage{ae,aecompl}

\usepackage{graphicx}	% Including figure files
\usepackage{amsmath}	% Advanced maths commands
\usepackage{amssymb}	% Extra maths symbols
\usepackage{natbib}
\usepackage{siunitx}    % for exponents like \num{1e-10} and degree symbol \ang{30}
\usepackage{float}
\usepackage{placeins}   % for keeping tables/figures where you want them
\defcitealias{Donnarumma2011}{D11}
\newcommand{\qlens}{\texttt{QLens}}

\title[Detecting Cores of Dark Matter Halos in Galaxy Clusters with Strong Lensing]{Detecting Dark Matter Cores in Galaxy Clusters with Strong Lensing}

\author[K. E. Andrade et al.]{
Kevin E. Andrade,$^{1}$\thanks{E-mail: kandrad1@uci.edu}
Quinn Minor,$^{2,3}$
Anna Nierenberg$^{4}$
and Manoj Kaplinghat$^{1}$
\\
$^{1}$University of California, Irvine, Irvine, CA 92697, USA\\
$^{2}$Department of Science, Borough of Manhattan Community College, City University of New York, New York, NY 10007, USA\\
$^{3}$Department of Astrophysics, American Museum of Natural History, New York, NY 10024, USA\\
$^{4}$Jet Propulsion Laboratories, Pasadena, CA 91109, USA
}

\pubyear{2018}

\begin{document}

\label{firstpage}
\pagerange{\pageref{firstpage}--\pageref{lastpage}}
\maketitle

\begin{abstract}
We test the ability of strong lensing data to constrain the size of a central core in the dark matter halos of galaxy clusters, using Abell 611 as a prototype. Using simulated data, we show that modeling a cluster halo with ellipticity in the gravitational potential can bias the inferred mass and concentration, which may bias the inferred central density when weak lensing or X-ray data are added. We also the highlight the possibility for spurious constraints on the core size if the radial density profile is different from the assumed model. These systematics can be ameliorated if central images are present in the data. Applying our methodology to Abell 611 and imposing a reasonable prior on the stellar mass-to-light ratio restricts the core size to be less than about 4 kpc, with a minimum reduced $\chi^2$ of 0.28 for $0\farcs2$ positional errors. Such small cores imply a constraint on the dark matter self-interaction cross section of the order of $0.1\ \mathrm{cm^2/g}$ at relative velocities of about 1500 km/s.
% Adding X-ray, weak lensing and stellar kinematics data sets will strengthen these constraints. 
% These results strongly argue that density profile measurements of a large sample of clusters will be an incisive tool for constraining dark matter self-interactions. 
% In addition, our results imply a bottom-heavy (super-Salpeter) initial mass function for the central galaxy in the Abell 611.
% These results indicate that with a careful treatment of systematics, strong lensing provides useful constraints on the density profile of dark matter in the central regions of galaxy clusters. 
\end{abstract}

% QM: I removed the following sentence from the abstract to shorten it...can put back in if desired?
%The finding of a small core is consistent with previous studies and is not significantly altered by stellar kinematic data. 


%%%%%%%%%%%%%%%%%%%%%%%%%%%%%%%%%%%%%%%%%%%%%%%%%%

%%%%%%%%%%%%%%%%% BODY OF PAPER %%%%%%%%%%%%%%%%%%

\section{Introduction}

Galaxy clusters provide a critical test of dark matter theories if their inner dark matter density profile can be measured. Hierarchical structure formation models make concrete predictions about cluster density profiles. For example, in the cold dark matter (CDM) paradigm, dark-matter-only simulations show that hierarchical structure formation leads to cuspy dark matter halos with a Navarro-Frenk-White (NFW) density profile \citep{Navarro1996,Navarro2010,Gao2012}, with the 3-D density profile $\rho \propto r^{-1}$ in the inner region. However, self-interactions or the feedback effects from baryons can potentially modify the inner slope. For example, Active Galactic Nuclei (AGN) may potentially cause flattening of central cusps of cluster mass halos \citep{Peirani2008,Martizzi2012,Mead2010}. Self-interacting dark matter (SIDM) models \citep{Spergel2000a} and simulations thereof \citep{Sokolenko2018a,elbert2015,rocha2013} predict shallower slopes for radial dark matter density profiles.

The presence or absence of dark matter cores in clusters is an open question. \citet{Sand2008} found an inner logarithmic slope of $\sim-0.5$ in two relaxed clusters using a combination of lensing and kinematic data. \citet{Newman2013a,Newman2013b} found a similar result for Abell 611 and similar clusters using a combination of strong lensing, weak lensing and stellar kinematics. \citet{DelPopolo2012,DelPopolo2014} came to the same conclusion analyzing a group of clusters that include Abell 611. In contrast, \citet{Caminha2017a} finds an inner log slope close to the canonical NFW value of -1 for the cluster MACS 1206. While simulations have been steadily advancing in scope and resolution, there is still no consensus in the question of cores in clusters (see \citealt{Schaller2015,Martizzi2012}).

The mass profiles of galaxy clusters can be probed by several methods, each having a distinct range of radii at which it can yield insight. These methods include stellar kinematics, strong lensing, X-ray emission and weak lensing, which cover the full range from 10 kpc to 1 Mpc scales~\citep{Umetsu2011,Newman2009,Hogan2017}.

% One in particular that we will discuss is well-known degeneracy between velocity anisotropy and mass \citep{Gerhard1993} in stellar kinematic analysis.
% However, each method has inherent systematic issues. Stellar kinematic analysis suffers from the well-known degeneracy between velocity anisotropy and mass \citep{Gerhard1993}. Weak lensing accuracy depends on the telescope and camera performance, and is affected by imperfections in the algorithm used in measuring galaxy shape \citep{Massey2013}. X-ray mass estimates of galaxy clusters have systematic biases from the different methods used to infer cluster masses, uncertainties in the mass-luminosity relation \citep{Rosati2002}, motions of the intracluster medium, and measurement of gas temperature \citep{Rasia2018}. 
% Moreover, how different measurements are combined is important, and methods for doing so are not standardized.

Strong lensing refers to multiple images of a background source, with the image positions and magnifications determined by the mass distribution of the deflector.  Multiple images can be exploited to provide strong constraints on the distribution of the matter in the lens, since the mass distribution must simultaneously satisfy the lens equations for all images \citep[][e.g.,]{Kneib2011}. Images appear near the Einstein radius of the lens, which is typically tens of kpc for galaxy cluster lenses, well within the cluster scale radius, and also where the effects of baryons and dark matter self-interactions are strongest. Separating the two  effects is critical.

Abell 611 is dynamically relaxed \citep[][hereafter D11]{Donnarumma2011}. It has been studied before, but usually in conjunction with X-ray images, weak lensing and/or kinematic data (see \citetalias{Donnarumma2011}, \citet{Schmidt2007, Romano2010, Richard2010, Newman2009, Newman2013a, Newman2013b,DelPopolo2012,DelPopolo2014, Monna2017,Zitrin2015}). Abell 611 is an excellent test case for gravitational lensing analysis, as it images multiple sources at multiple redshifts, contains radial arcs at various locations, and has images at a range of radii, from 30 kpc to more than 100 kpc. 

In this work we constrain the dark matter distribution using strong lensing alone. This allows us to characterize the strengths and limitations of strong lensing separately from other techniques. Our primary goal is to determine how well one can constrain the size of a central core in a lensing cluster, i.e., determine the radius (if any) below which the density profile becomes relatively flat.

We adopt a flat cosmology with $\Omega_{\Lambda}=0.7$, $\Omega_m=0.3$, and $H_0=70\: km\: s^{-1} Mpc^{-1}$. At the adopted lens redshift of 0.288, the distance to the lens is 893.8 Mpc, and 1\arcsec is equal to 4.329 kpc. We define halo mass as $M_{200}$, the mass enclosed by a sphere of radius $R_{200}$, which we define in turn as the radius at which the halo density is 200 times the critical density of the universe at the redshift of the halo.

This paper is organized as follows. In Section~\ref{sec:Lensing Software} we discuss the new lensing code and lens mass profiles. In Section~\ref{sec:Mock Data Modeling} we discuss the mock data sets and lens models used in our analysis, and the results of those models. In Section~\ref{sec:Modeling of Abell 611} we describe the data and lens model for Abell 611, and discuss the results of the analysis for that cluster. Our conclusions are summarized in Section~\ref{Conclusions}.

\section{Halo Models and Lensing Software} 
\label{sec:Lensing Software} 

We choose a flexible mass model for our tests which has a core, and for which the mass distribution approaches NFW as the core radius goes to zero. We require a fast method to calculate the magnification and deflection at each point. These requirements are in a new lensing software, \qlens, which allows for both pixel image modeling (using pixelated source reconstruction) and point image modeling (with option to include fluxes, time delays, and multiple sources at different redshifts). \qlens~includes 14 different analytic lens models to use for model fitting, including 10 different density profiles where ellipticity can be introduced into either the projected density or the lensing potential. In addition, a built-in nested sampler is included, along with an adaptive MCMC algorithm called T-Walk; however the code can also be compiled with MultiNest or PolyChord. The \qlens~package is now available on GitHub by request and includes a student-friendly tutorial to get users started with point image modeling. Here, we describe the novel features implemented in \qlens~that are critical to the results of this paper.

We consider two types of cored halo models for modeling cluster halos. The first model is a cored NFW profile (cNFW), for which the density profile is defined as

\begin{equation}
\rho = \frac{\rho_s r_s^3}{\left(r_c + r\right)\left(r_s + r\right)^2}.
\end{equation}

This will be the primary lens model we use in this work for the cluster halo. Note that as the core radius $r_c \rightarrow 0$, the density profile reduces to the standard NFW form. This profile has been used in lens modeling in \citet{Newman2013a,Newman2013b}, and was found by \cite{penarrubia2013} to provide a reasonably good fit to cored DM halos found in hydrodynamical simulations of \cite{governato2012}.  Analytic formulas for the kappa profile and deflection of the corresponding spherical model are given in Appendix \ref{sec:appendix_a}.

The second cored halo model, which we call the \emph{Corecusp} model, is defined as

\begin{equation}
\rho = \frac{\rho_0 r_s^n}{\left(r^2+r_c^2\right)^{\gamma/2} \left(r^2+r_s^2\right)^{(n-\gamma)/2}}.
\end{equation}

This is an extension of the ``cuspy halo model'' of \cite{munoz2001}. This model also allows for a core in addition to a scale radius, where $r_c < r_s$. In the limit of large $r$, the log-slope is given by $n$, whereas in the limit of small $r$ and zero core, the log-slope is given by $\gamma$. Note that in this model, both the core and scale radius are added in quadrature to $r$, resulting in a more rapid turnover compared to the cNFW model at the scale and core radii.  As a result, the profile does not reduce exactly to NFW in the limit $r_c \rightarrow 0$. On the other hand, the greater flexibility afforded by the variable inner and outer log-slopes may become useful when combining strong lensing with data that probe the density profile on larger scales, e.g. weak lensing or X-ray data. For the purposes of this paper, however, this profile 
provides a comparison of how sensitive the core constraints are to the exact nature of the turnover behavior of the density profile near $r_c$. As we are primarily interested in the behavior in the region interior to $r_s$, we fix n to 3 in this work, to match that of an NFW profile. The relevant lensing formulas are given in Appendix \ref{sec:appendix_a}.

For each of the above lens models, we add ellipticity by making the replacement $R^2 \rightarrow qx^2 + y^2/q$ in the projected density profile. The deflection and Hessian of the lens mapping must be calculated by numerical integration (see \citealt{schramm1990,keeton2001b}), which is computationally expensive. While the integrals can be done relatively quickly using Gaussian quadrature, it is not known \emph{a priori} how many points will be required for the integral to converge beyond a specified tolerance. This can be solved by an adaptive quadrature scheme where the integration is done at successively higher orders and an error estimate is obtained after each iteration, then stopping when the error falls below a specified tolerance.  To implement this, we employ a modification of Gaussian quadrature known as Gauss-Patterson quadrature, which consists of nested quadrature rules whereby a given order of integration retains the function evaluations from the lower orders, thus ensuring they are not wasted (at the cost of allowing up to a maximum order of 511 points).\footnote{The algorithm described above is quite similar to adaptive Clenshaw-Curtis quadrature except it is an open interval quadrature rule, thus dodging the issue of having to evaluate the projected density or its derivative at $r=0$.} For lensing calculations, we find this adaptive quadrature scheme requires nearly an order of magnitude fewer function evaluations compared to Romberg integration for a tolerance $\sim 10^{-3}$. This reduces the expense of lensing calculations enormously for elliptical projected density profiles.

The mass distribution of the dark matter halo in strong lenses has been shown to be consistent with elliptical isodensity contours in several studies \citep{yoo2005,yoo2006,kochanek2004}. Hence, when modeling the cluster halo, there is strong motivation for introducing ellipticity into the projected density profile as described above. However, because of the computational burden of performing the integrations for elliptical density profiles, it is common to instead use the ``pseudo-elliptical'' model in which the halo ellipticity is incorporated into the gravitational potential rather than the projected density \citep{Golse2002}. Here, we consider both approaches, and will compare the pseudo-elliptical approximation to the full elliptical density approach.

\section{Mock Data Modeling}
\label{sec:Mock Data Modeling}

\begin{table*}
	\centering
    \caption{The parameters, their true values, prior ranges and prior types for the mock data set. For all free parameters, the prior distribution is uniform over the parameter range.}
	\label{tab:mock data parameters}
	\begin{tabular}{llcccc} % columns, alignment for each
		\hline
		Parameter Name & Description & Units & True Value & Prior Range \\
		\hline
        \textit{DM Halo (cored NFW profile)}\\
		$M_{200}$ & halo mass & $M_{\sun}$ & $\num{1.1e15}$ & $\num{2e14} - \num{5e15}$ \\
        $c_{200}$ & concentration & - & 7.0 & 1 - 20\\
        $\beta_c$ & core ratio ($r_c/r_s)$ & - & 0.157 & 0.0 - 0.96 \\
        q & axis ratio & - & 0.8 & 0.3 - 1.0 \\
        $\theta$ & orientation & degrees & 132.5 & 120 - 150 \\
        x-center & x coordinate of center &\arcsec & 0.& -5 - 5 \\
        y-center & y coordinate of center &\arcsec & 0.& -5 - 5 \\
        \\
        \textit{BCG (dPIE profile)}\\
        b & mass parameter &\arcsec & 0.60& 0.1 - 10.0 \\
        a & scale radius &\arcsec & 15& (fixed) \\
        s & core radius &\arcsec & 0.05& (fixed) \\
        q & axis ratio & - & 0.75& (fixed) \\
        $\theta$ & orientation & degrees & 72.5& (fixed) \\
        x-center & x coordinate of center &\arcsec & 0.5& (fixed)\\
        y-center & y coordinate of center &\arcsec & -0.9& (fixed)\\ 
		\hline
	\end{tabular}
\end{table*}

\begin{table*}
	\centering
    \caption{Mock data calculated parameters.}
	\label{tab:mock data calculated parameters}
	\begin{tabular}{llccc} % columns, alignment for each
		\hline
        Parameter Name & Description & Cored System & Cuspy System\\
        \hline
        $D_{lens}$ & angular dia. dist. to lens & 894 Mpc & 894 Mpc\\
		$\Sigma_{crit} (z_{src, ref}=1.49)$ & critical surface density & $\num{2.53e9} M_{\sun}/kpc^2$ & $\num{2.53e9} M_{\sun}/kpc^2$\\
        \\
        \textit{  DM Halo (cNFW profile)}\\
        $r_s$ & scale radius & $ 63\farcs7$ & $63\farcs7$  \\
        $r_c$ & core radius & $10\arcsec$ & $0\arcsec$\\
        $r_{200}$ & halo radius & $445\farcs7$ & $445\farcs7$\\

        \\
        \textit{  BCG (dPIE profile)}\\
        $r_{E, BCG}$ & Einstein radius & $0\farcs57$ & $0\farcs57$ \\
        $M_{BCG}$ & total mass & $\num{1.34e+12} M_{\sun}$ & $\num{1.34e+12} M_{\sun}$\\
		\hline
	\end{tabular}
\end{table*}

We are interested in determining the capabilities of strong lens modeling for inferring the cluster dark matter halo properties. The following questions guided our choice of mock data sets.
\begin{enumerate}
\item Is it possible to distinguish between a core and a cusp with strong lensing alone?
\item To what extent can central images help in determining the inner density profile?
\item How do inaccuracies in the outer density profile affect the result? Do they lead to a spurious detection of a core or cusp?
\item Does the use of an elliptical \textit{potential} rather than using a elliptical \textit{density} profile lead to inaccurate results? 
\end{enumerate}
Modeling the mock data allows us to test the power of models to constrain relevant parameters in a controlled way. 

\subsection{Mock Data Preparation}
\label{sec:Mock Data Preparation}

\begin{figure*}
      \centering
       \caption{The location of the mock data image points, with representative caustic curves and critical curves. The representative curves shown correspond to a reference redshift of $z=1.49$.  [\textit{Top Row}] Source plane plots for the cases that exclude central images. Cored is left, cuspy  is right.  [\textit{Bottom Row}] Image plane plots for the base cored case (left) and base cuspy case (right).}
      \includegraphics[width=0.49\textwidth]{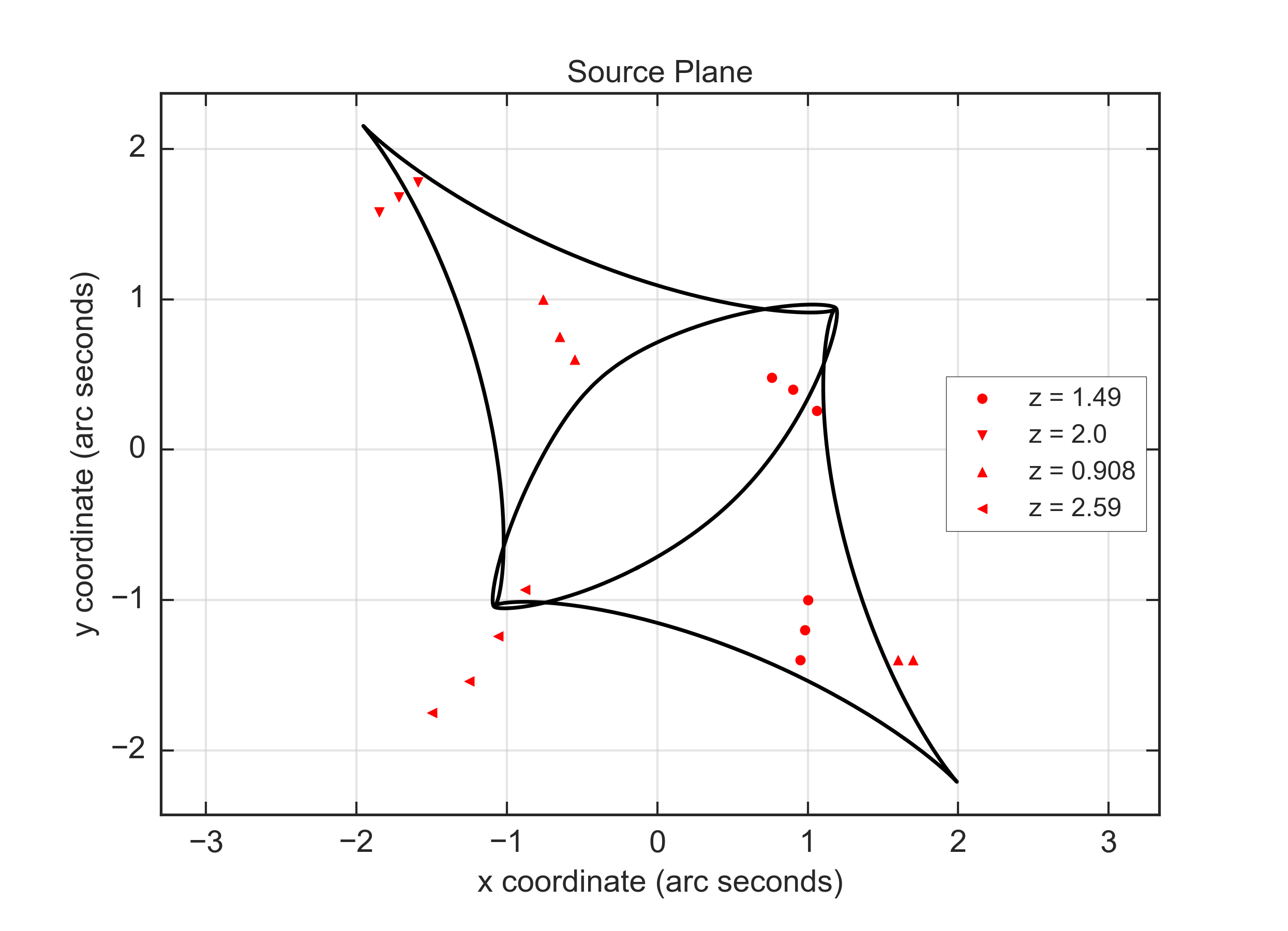}
      \hfill
      \includegraphics[width=0.49\textwidth]{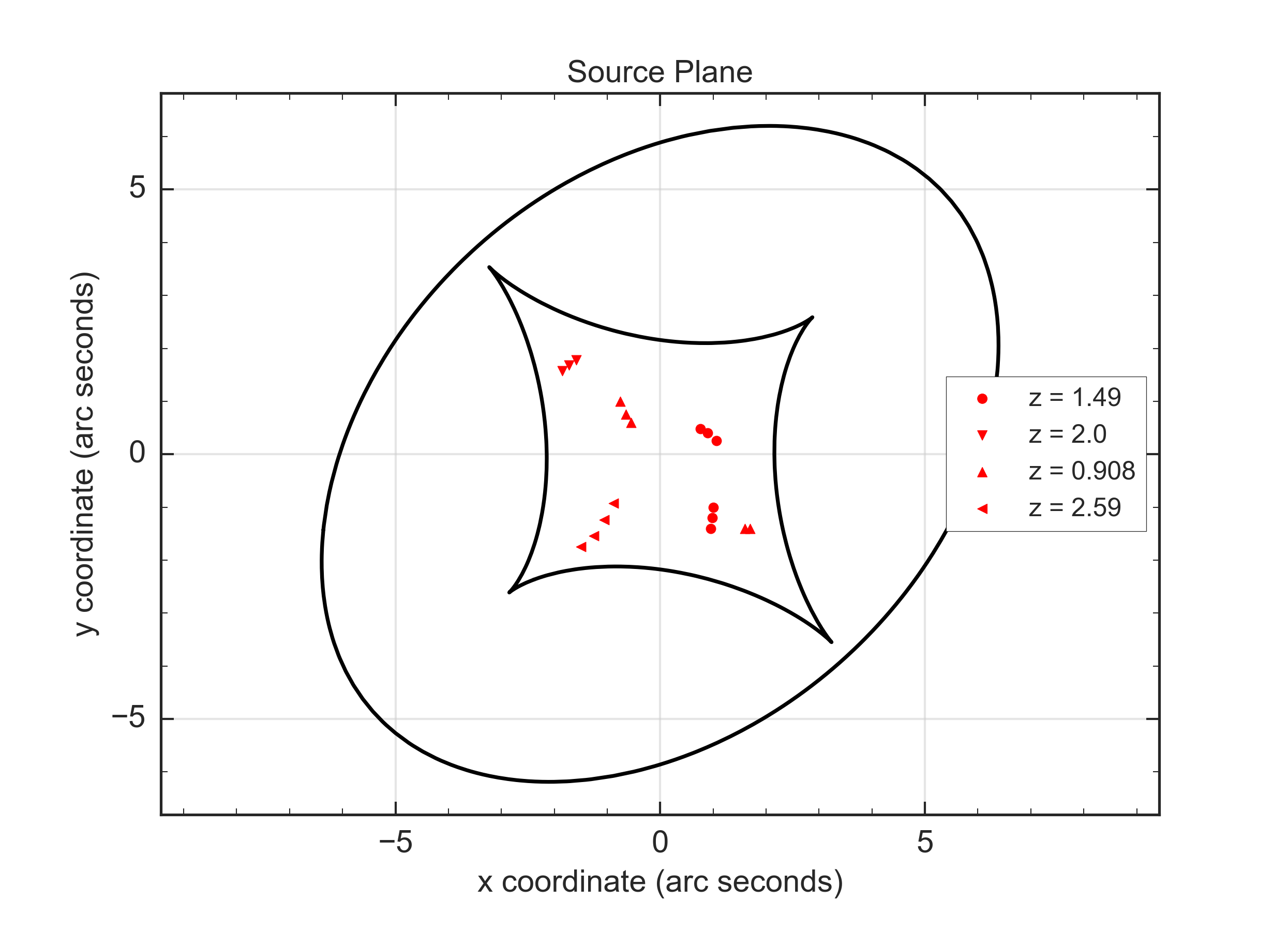}\\
      \includegraphics[width=0.49\textwidth]{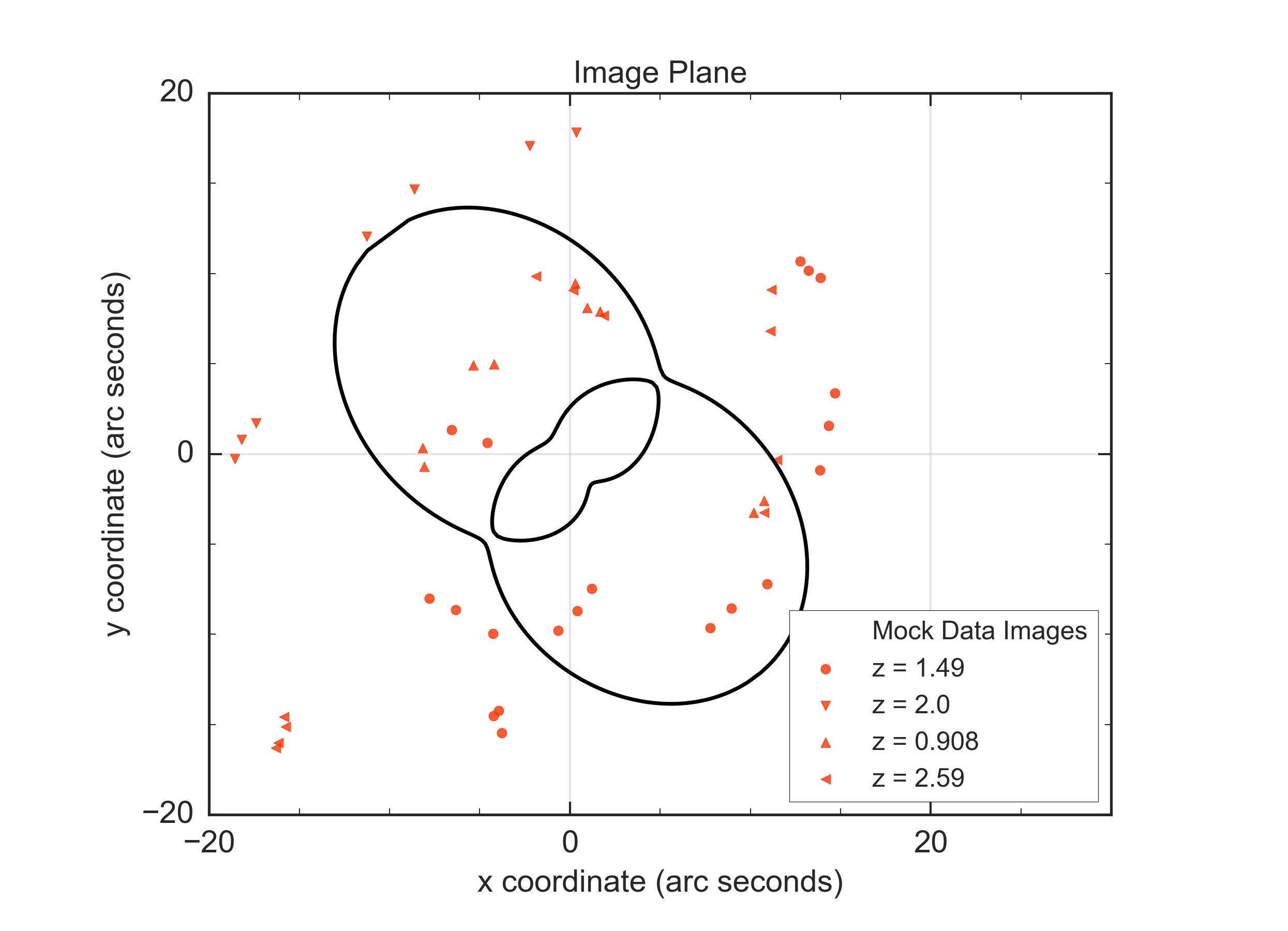}
      \hfill
      \includegraphics[width=0.49\textwidth]{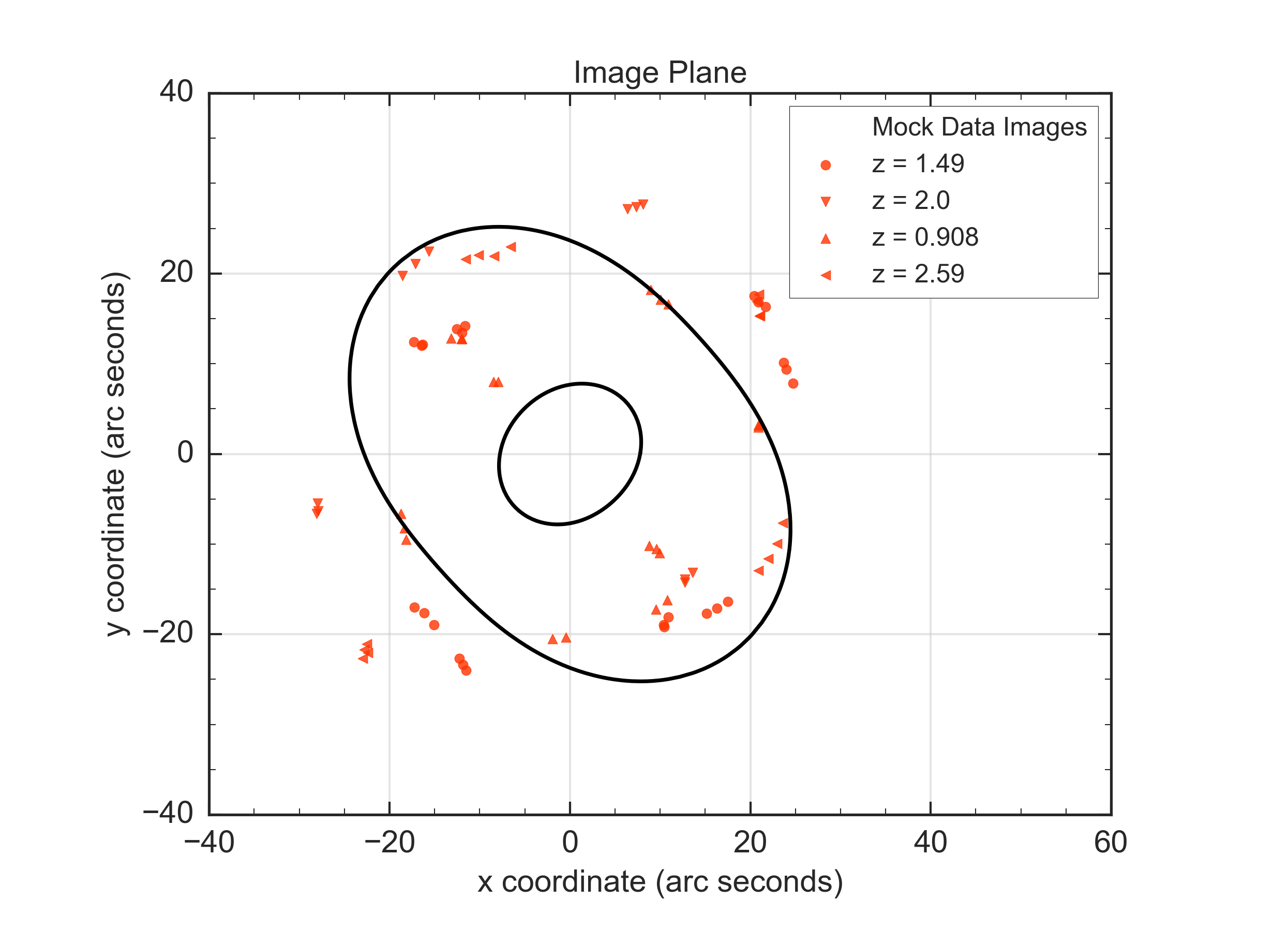}\\
      \label{fig:mock_data}
\end{figure*}

To test the ability of the lens models and software to constrain the relevant halo and BCG parameters six sets of mock data were created, as follows:
\begin{itemize}
\item Cuspy, no central images
\item Cored, no central images
\item Cuspy, with central images
\item Cored, with central images
\item Cored, no central images, highly elliptical halo (axis ratio = 0.5) 
\item Cuspy, no central images, highly elliptical halo (axis ratio = 0.5) 
\end{itemize}

The mock data sets were constructed to be similar in nature to Abell 611 in most respects, including mass, redshift, position angle, offset and ellipticity. To examine the usefulness of central images in constraining system parameters, two image sets were generated for each of the cored and cuspy cases; one image set included positive parity central images and one did not. The input parameters of the mock data objects are summarized in Table~\ref{tab:mock data parameters}, and calculated parameters are shown in Table~\ref{tab:mock data calculated parameters}. 

The mock data sets consisted of a dark matter halo and a bright central galaxy, offset by $\sim1$\arcsec, each at a redshift of 0.288, matching the redshift and inferred offset of Abell 611 \citepalias{Donnarumma2011}. The dark matter halo was modeled by a cNFW profile. The scale radius (``$r_s$") was chosen to be 50\arcsec and the halo mass to be $\num{1.1e15}M_{\sun}$, similar in magnitude to Abell 611 and other galaxy clusters. The dark matter halo is oriented \ang{132.5} counterclockwise from the x-axis. In the cored cases only, a uniform density core is modeled with a transition radius of 10\arcsec.

We chose an axis ratio of 0.8 for primary mock data sets, as that is similar to that of typical clusters. But since galaxy clusters can sometimes have highly elliptical structure (see \citet{Richard2010}, Table 7), we constructed two separate mock data sets with a highly elliptical dark matter halo (axis ratio = 0.5) to investigate the effects of more severe ellipticity on the model inferences, which we discuss in Section \ref{Pseudo-Elliptical Approximation}. 

The BCG was modeled with a dual pseudo-isothermal ellipsoid (dPIE) profile (defined in Appendix \ref{sec:lensing_formulae}) that allows separate specification of the tidal break radius and core radius. The \qlens~mass parameter for dPIE profiles,``b", can be expressed as
\begin{equation}
\label{eq: b}
	b=\frac{\sigma_0^2 r_{cut}}{2 G \Sigma_{crit}(r_{cut}-r_c)}
\end{equation}
where $\sigma_0$ is the central velocity dispersion, G is the gravitational constant, $\Sigma_{crit}$ is the critical surface density of the lens at the relevant redshift, $r_{cut}$ is the tidal cutoff radius and $r_c$ is the core radius. Therefore, the mass parameter ``b" roughly corresponds to the Einstein radius (and reduces exactly to the Einstein radius as $r_c\rightarrow0$), and $b\propto\sigma_0^2$. The BCG was given a stellar mass of $\num{1.34e12}M_{\sun}$, which corresponds to $b=0.60$, and a small core of uniform density, with a radius of 0.05\arcsec, similar to that noted in Abell 611 \citepalias{Donnarumma2011}. 

The tidal break radius of the BCG in Abell 611 found in recent literature is between 10 and 20\arcsec (\citet{Newman2009} and \citepalias{Donnarumma2011}); a value of 15\arcsec was chosen for this mock data analysis. The axis ratio was chosen at 0.75, oriented at an angle of \ang{72.5}, which is \ang{60} clockwise from the halo orientation. The BGC was positioned at ($0\farcs5$, $-0\farcs9$) in the image plane. This corresponds to an offset from the dark matter halo by $\sim4$ kpc, consistent with the offset found for Abell 611 in \citetalias{Donnarumma2011} and \citet{Newman2009}, and similar to that of other clusters \citep{Newman2013a}.

Sources for the mock data were chosen in four redshift groups: 0.908, 2.00, 2.06 and 2.59, respectively, similar to source redshifts Abell 611 \citepalias{Donnarumma2011}. For each redshift, one or two compact sources were created within an area of approximately 1\arcsec, with two to three source points each. Simulated position errors with a standard deviation of $0\farcs2$ were incorporated into the image positions. Figure~\ref{fig:mock_data} shows the source plane and image plane representations for the cored and cuspy data sets. 

\subsection{Mock Data Lens Model}
\label{sec:Mock Data Lens Model}

A lens system model was constructed with two lens objects, one to represent the halo and one to represent the BCG. As in the mock data preparation, the halo lens was modeled with a cNFW profile and the BCG lens with a dPIE profile. 

\begin{figure}
    \caption{Example of a final fit image for the mock data. This fit is for the cuspy halo, without central images. The data points are shown in red, and the modeled images in cyan. The points appear as purple when the best fit model and data images overlay. The unmatched model images near (0, 0) are positive-parity ``central" images, which are typically unobservable due to low magnification and/or obscuration by bright objects in the center of the cluster.}
	\includegraphics[width=\columnwidth]{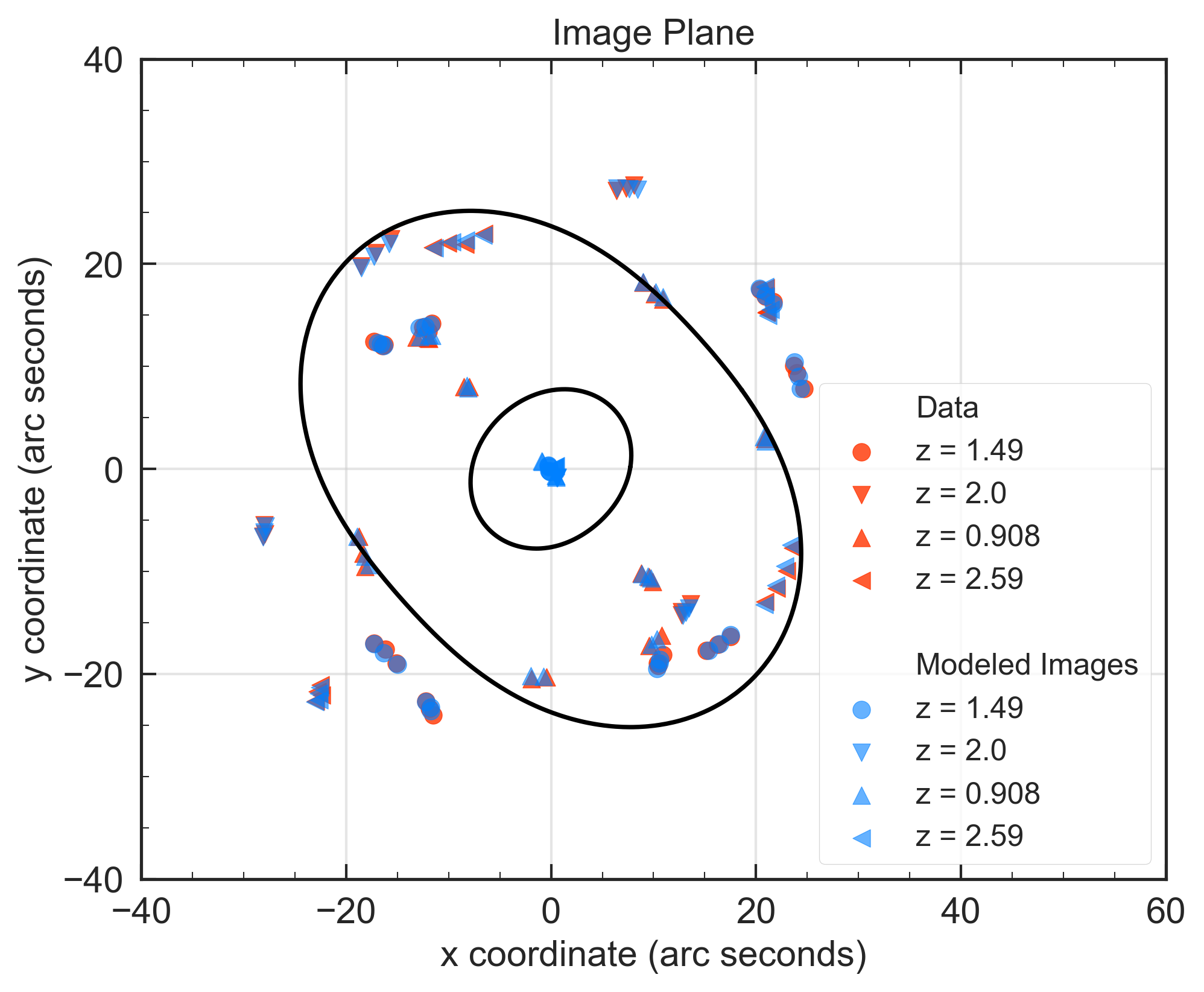}
    \label{fig:fit_example}
\end{figure}

\begin{figure}
    \caption{Core radius posterior distributions for the cored cases (upper) and cuspy cases (lower). The true value for core radius is 43.3 kpc (10\arcsec) in the cored case and zero in the cuspy case. Note the different horizontal scales.}
	\includegraphics[width=\columnwidth]{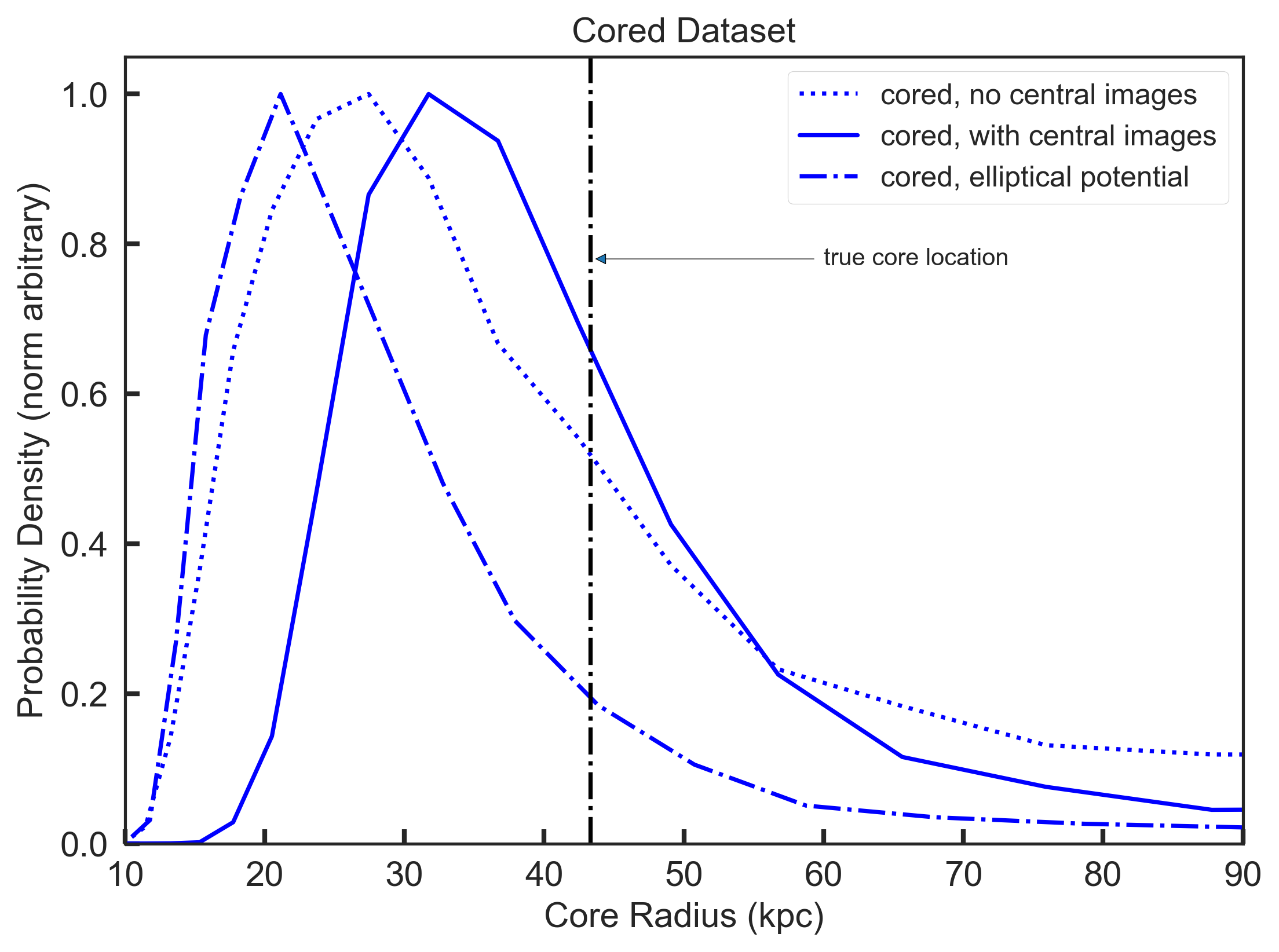}\\
    \includegraphics[width=\columnwidth]{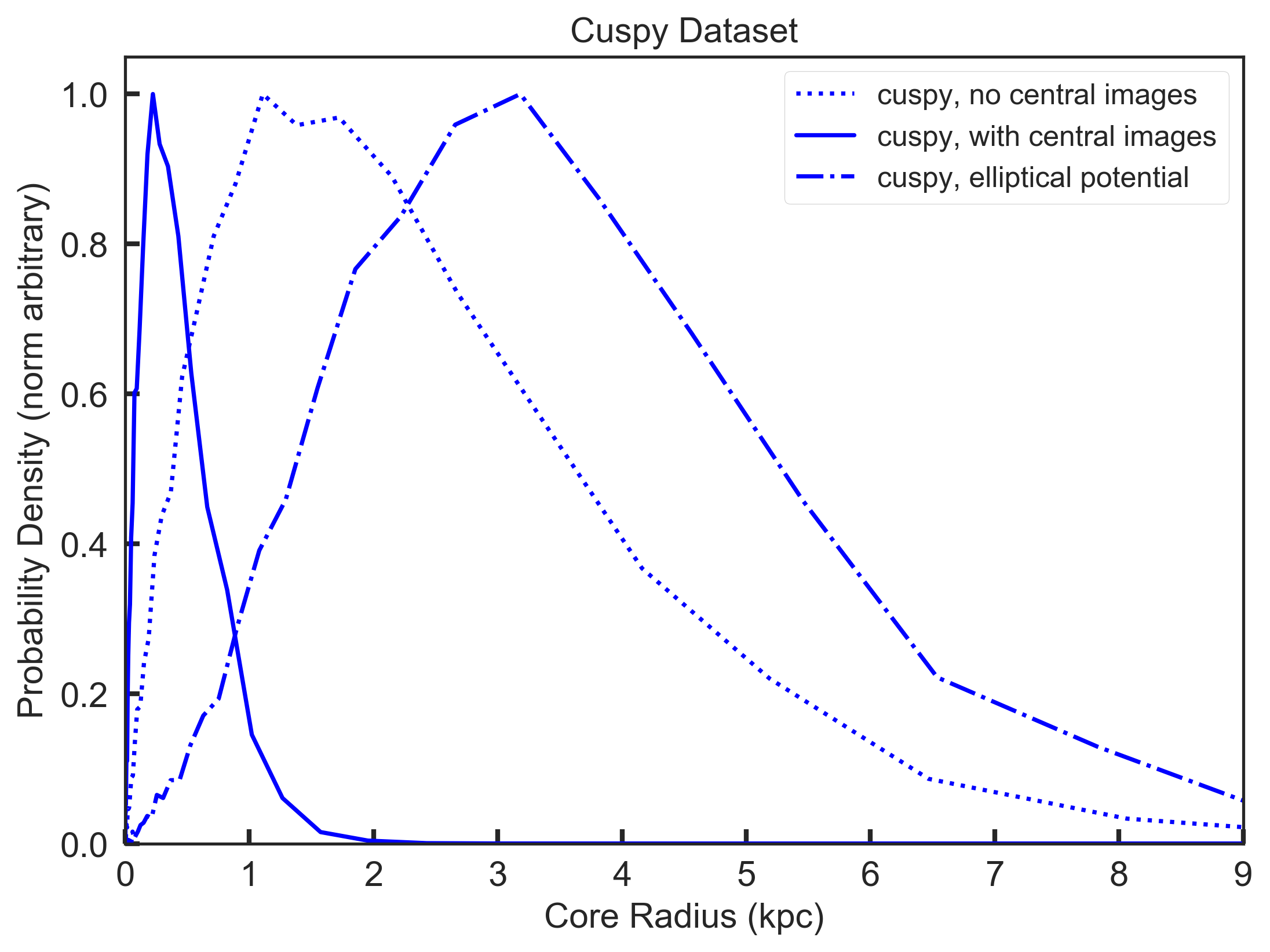}\\
    \label{fig:core_radius}
\end{figure}

The systems were analyzed using a nested sampling algorithm with 1,000 live points. There are 8 free parameters (7 for the dark matter halo and one for the BCG). Source plane $\chi^2$ evaluations were used in the beginning of each run, with a switch to image plane evaluations occurring mid-run. The image plane $\chi^2$ is calculated as follows:

\begin{equation}
\chi^2_{img}=\sum_{i} \frac{\left(\mathbf{x}_{obs,i}-\mathbf{x}_{mod,i}\right)^2}{\sigma_i^2}
\end{equation}
where i is the image index, $\sigma_i$ are the image position uncertainties, $\mathbf{x}_{obs,i}$ is the observed image position and $\mathbf{x}_{mod,i}$ is the modeled image position.
A similar $\chi^2$ can be calculated in the source plane, as described in \citet{Keeton2001}.

We used uniform priors for the parameters as shown in Table~\ref{tab:mock data parameters}. All of the models showed a close match between the data images and the model images. An example final fit image is shown in Figure~\ref{fig:fit_example}. 

\subsection{Mock Data Results}
\label{Mock Data Results}

\subsubsection{Data without Central Images}
\label{Data without Central Images}
We first consider as our baseline a dataset with no central images included, as might be expected for a cluster system with a bright object near the center that would obscure such central images. Figure~\ref{fig:core_radius} shows the posterior probability profiles for the cNFW ``$r_c$'' parameter (core break radius) for the six cases. For the cored data set with no central images (dotted curve), the median value of core radius (true value of 10\farcs0) is 7\farcs0, with a 1-$\sigma$ lower bound of 4\farcs8. For the cuspy data set (true core radius of zero), the median fit value is 0\farcs3, with a 1-$\sigma$ upper bound of 0\farcs6. The model is able to accurately distinguish between the cored and cuspy cases.

Triangle plots showing the posterior distributions of the parameters can be found in Figures~\ref{fig:cored_triangle} and \ref{fig:cuspy_triangle} in Appendix~\ref{triangle_plots}. The parameters are successfully recovered, with all of the true values of the parameters within $\sim 1$ standard deviation of the best-fit posterior value. 

\begin{figure}
      \centering
       \caption{Plots of scaled surface density ($\kappa$) versus radius for the mock data models. The use of the central images in the fitting enables a tighter constraint to the mass density in the inner region. The median (50th percentile) posterior value of the parameter set is shown as a solid red line, and the 16- to 84-percentile band is shown in gray. The true parameter value is shown as a dashed blue line. The radii bands in which images are located are highlighted in red. A reference redshift of $z=1.49$ is used in the calculation of $\kappa$. The plots are slices that are averaged over $\ang{360}$. The BCG can be observed as the bump at a radius of approximately $1\arcsec$. } [\textit{From top}]: Cored without central images, cupsy without central images, cored with central images, cuspy with central images.
       % format for trim is [trim=left bottom right top, clip]
      \includegraphics[trim=10 53 10 5, clip, width=\columnwidth]{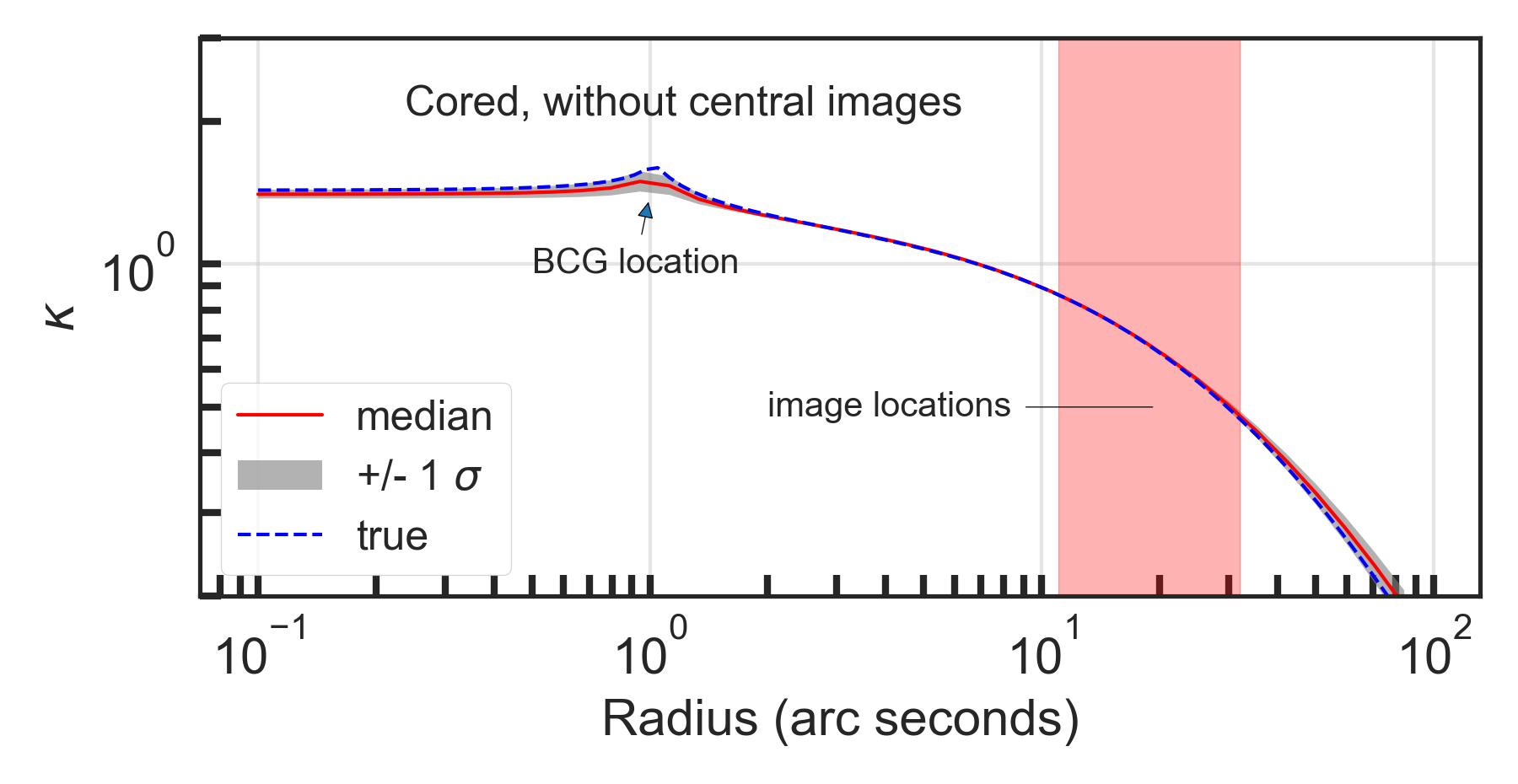}\\
      \includegraphics[trim=10 53 10 11, clip, width=\columnwidth]{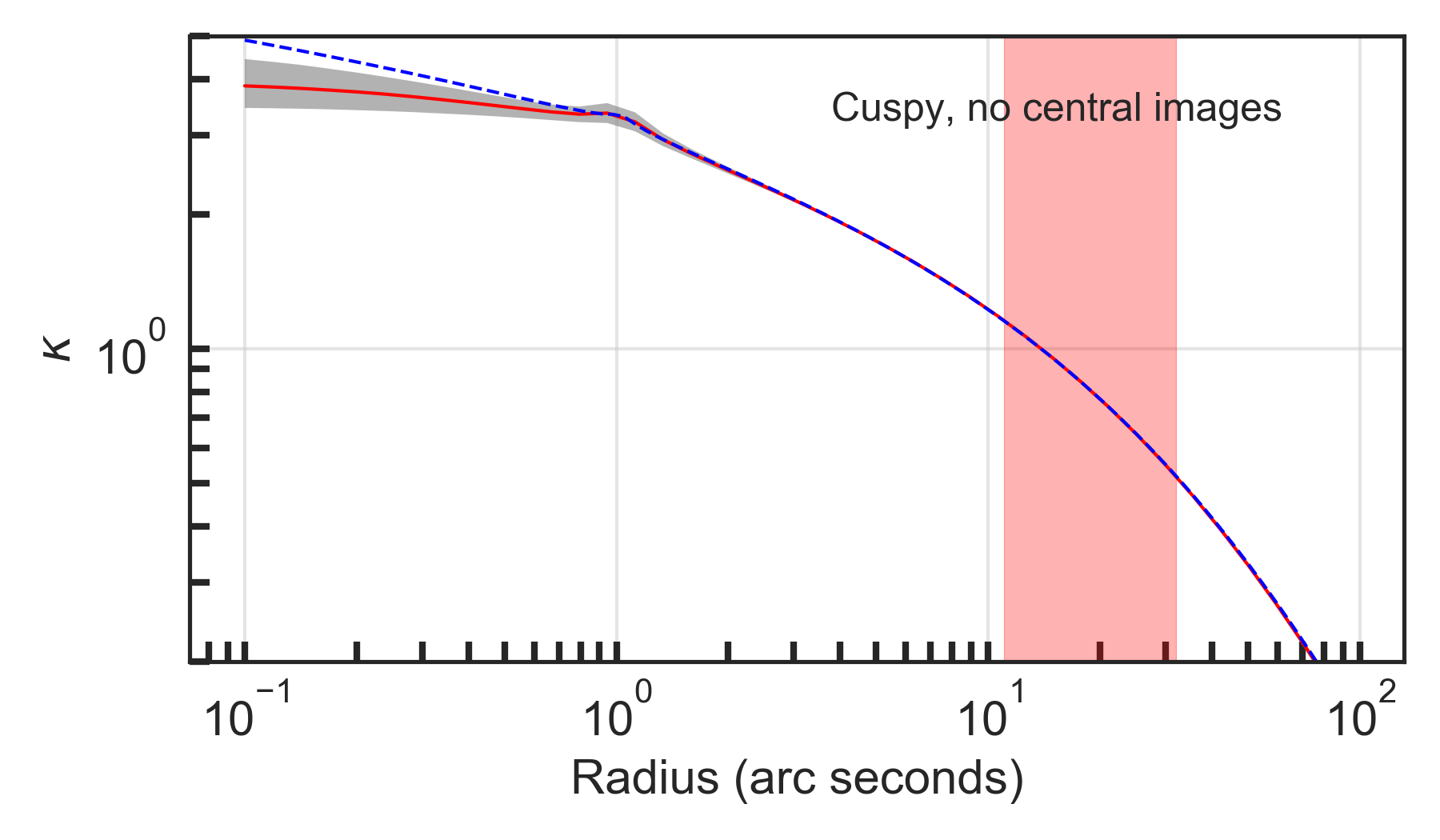}\\
      \includegraphics[trim=10 53 10 11, clip, width=\columnwidth]{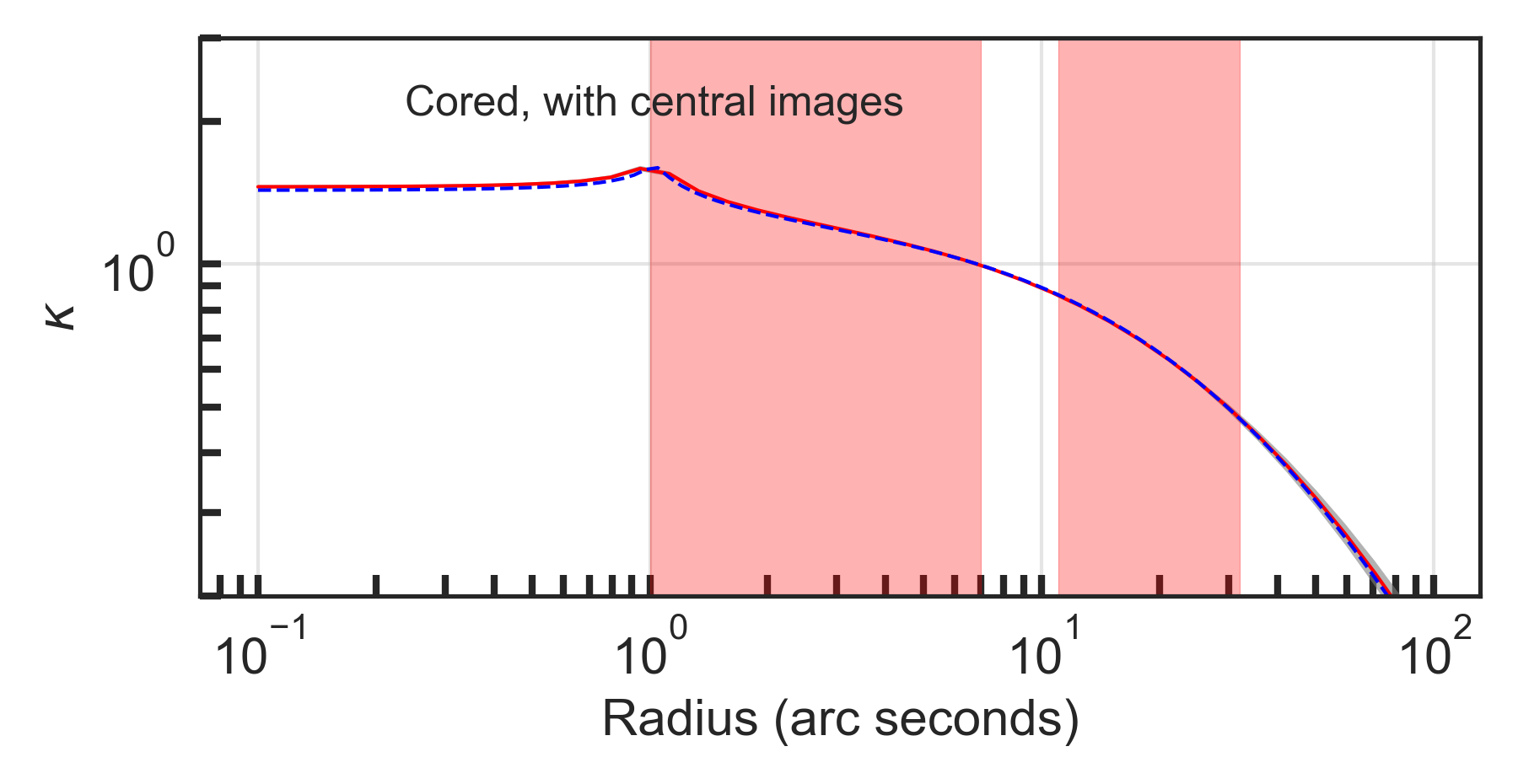}\\
      \includegraphics[trim=10 0 10 11, clip, width=\columnwidth]{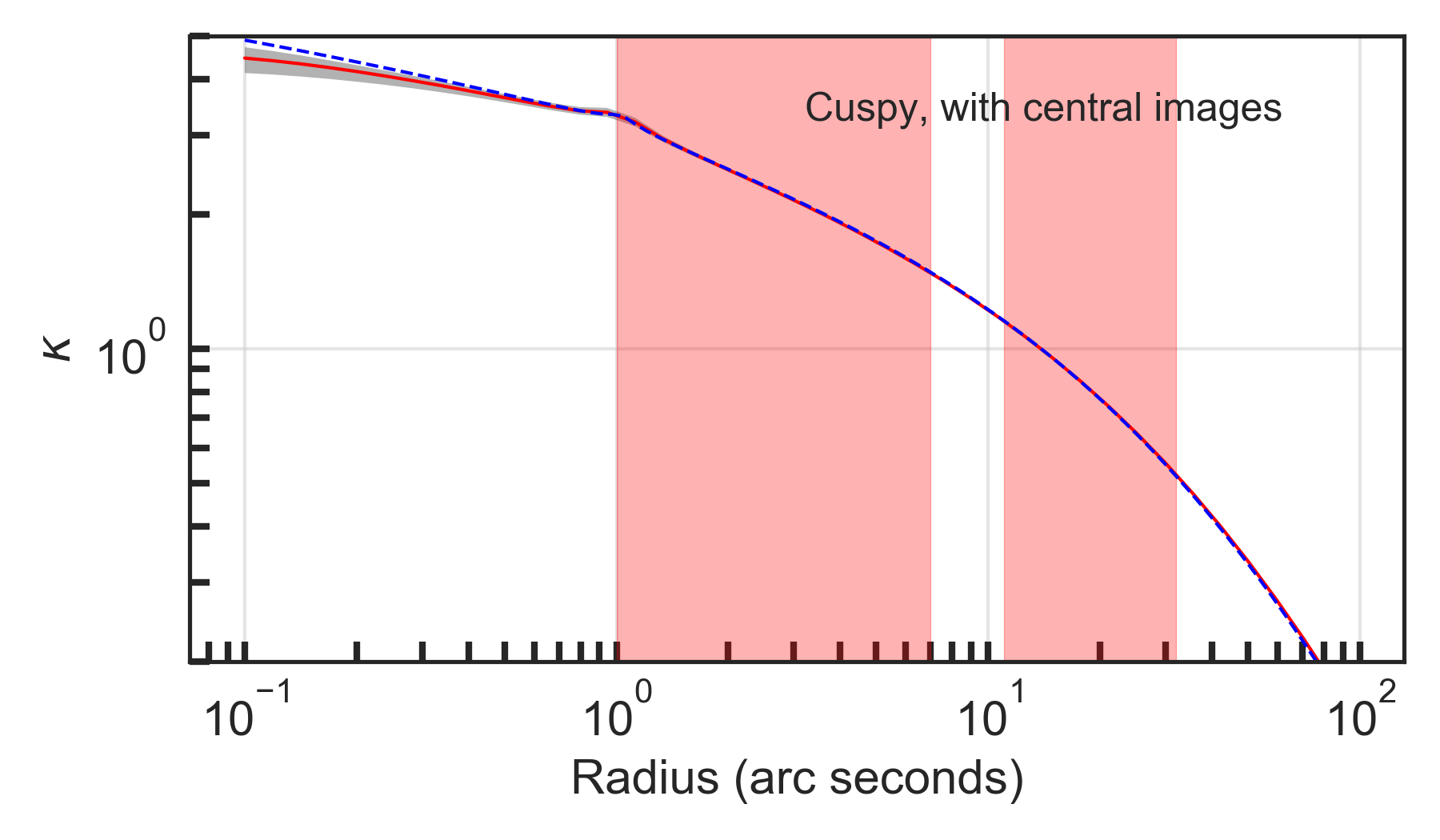}\\
      \label{fig:kappa}
\end{figure}

\subsubsection{Data with Central Images}
\label{Data with Central Images}
It is interesting to look at the same analysis with the central images included to see how useful the central images are in constraining parameter values. The solid curves of Figure~\ref{fig:core_radius} provide an illustration of this. For the cored data set, the best fit value of core radius is 8\farcs5, with a 1-$\sigma$ lower bound of 6\farcs5. For the cuspy data set, the best fit value is 0\farcs05, with a 1-$\sigma$ upper bound of 0\farcs12. Clearly, the central images greatly enhance the ability of the model to accurately constrain the core radius. The other parameters follow a similar pattern, with the mass and scale radius parameters of the dark matter halo determined with some uncertainty (but with more certainty than in the cases without central images), while the axis ratio, position angle and centroid coordinates are determined with high certainty.

\subsubsection{Constraints on Surface Density}
\label{Constraints on Surface Density}
One measure of the utility of the model is the ability to accurately reproduce the (2-dimensional) surface density of the cluster. Here we examine scaled surface density, $\kappa(r) \equiv \frac{\Sigma(r)}{\Sigma_{crit}}$, where $\Sigma(r)$ is the surface density and $\Sigma_{crit}$ is the critical surface density for the pertinent lens and source redshifts. Figure~\ref{fig:kappa} shows the circularly-averaged $\kappa$ versus radius for the cored and cuspy data sets, both with and without central images. Surface density is very accurately determined in the region where images are present, and the presence of central images enhances the accuracy of the predictions in the inner regions. Only in the radial regions far from the images does the predicted $\kappa$ deviate significantly from its true value.

\subsection{Pseudo-Elliptical Approximation}
\label{Pseudo-Elliptical Approximation}
As discussed in Section \ref{sec:Lensing Software}, an often-employed approximation is to use an elliptical form for the gravitational potential of the object rather than the density itself \citep{Golse2002}. The limits of validity for that approximation is given in \citet{Golse2002} to be for the range of ellipticities $\epsilon\lesssim0.25$, which corresponds to an axis ratio $q\gtrsim0.75$. Here we compare the results of such an approximation to that of using the true elliptical density. The dot-dashed lines in Figure~\ref{fig:core_radius}  show the cored and cuspy cases (without central images) but using the pseudo-elliptical approximation, for the mock data set with axis ratio $q=0.8$. The pseudo-elliptical model has somewhat less power to resolve the different cases than the model that uses the full elliptical density (solid lines in the figure). Specifically, in the cored case, where the true value of the core radius is 43.3 kpc, the core radius posterior for pseudo-elliptical model peaks at $\sim$20 kpc, while the true elliptical model shows a peak at $\sim$30 kpc. In the cuspy case, where the true core radius is zero, the pseudo-elliptical model produces a peak posterior at $\sim$3 kpc, whereas the true elliptical model has a peak at $\sim$1 kpc.

\begin{figure*}
	\caption{Posterior distributions for halo mass, concentration and core radius parameters, showing the effect of using an approximate elliptical potential for highly elliptical systems. [\textit{Top Row:}] full elliptical density profile used. [\textit{Bottom Row:}] pseudo-elliptical approximation used. [\textit{Left Column:}] cored case. [\textit{Right Column:}] cuspy case. Orange lines and 'x' markers indicate the true parameter values. The units for $m_{vir}$ are $10^{15}M_{\sun}$, and for core radius, arcseconds.}
	\includegraphics[width=\columnwidth]{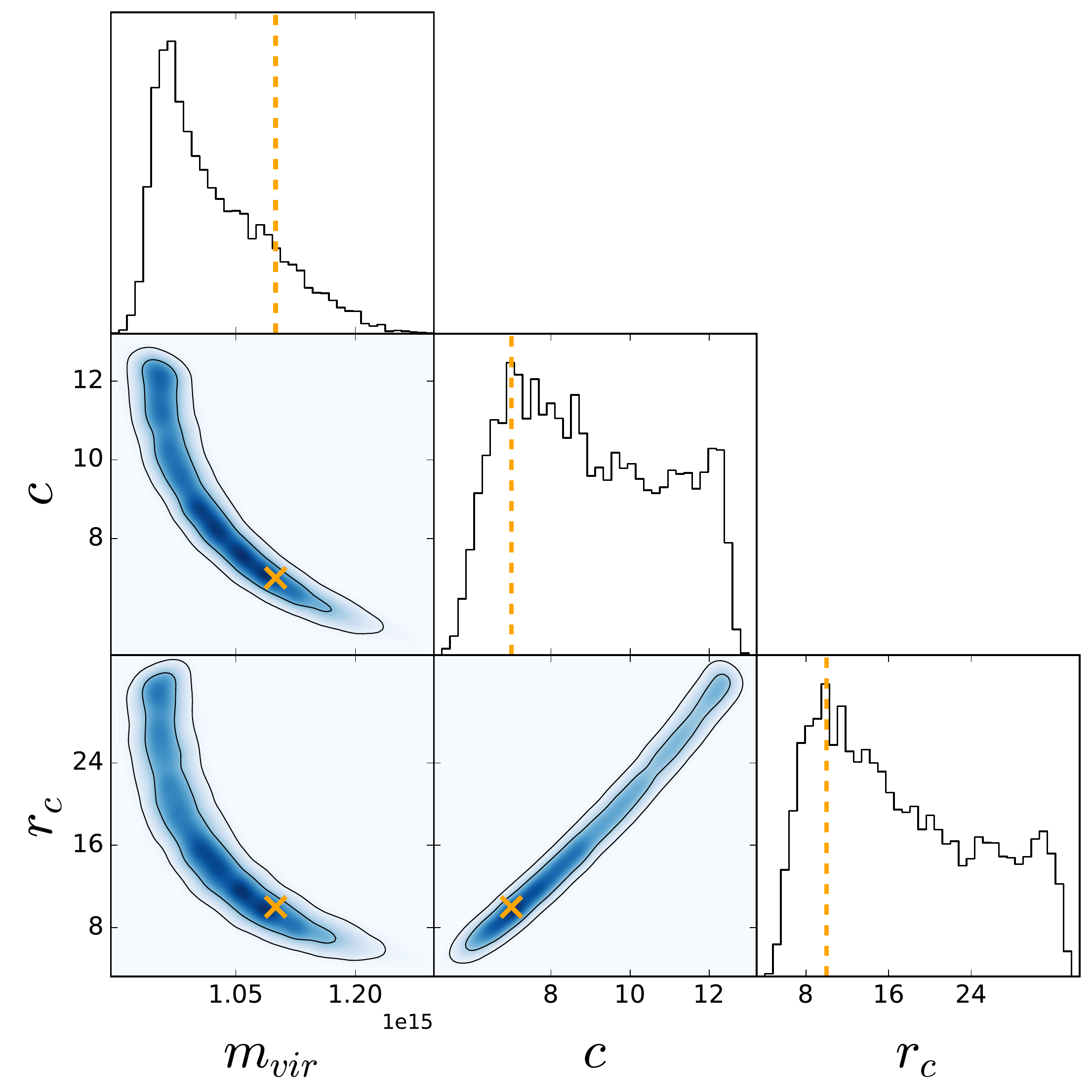}
    \includegraphics[width=\columnwidth]{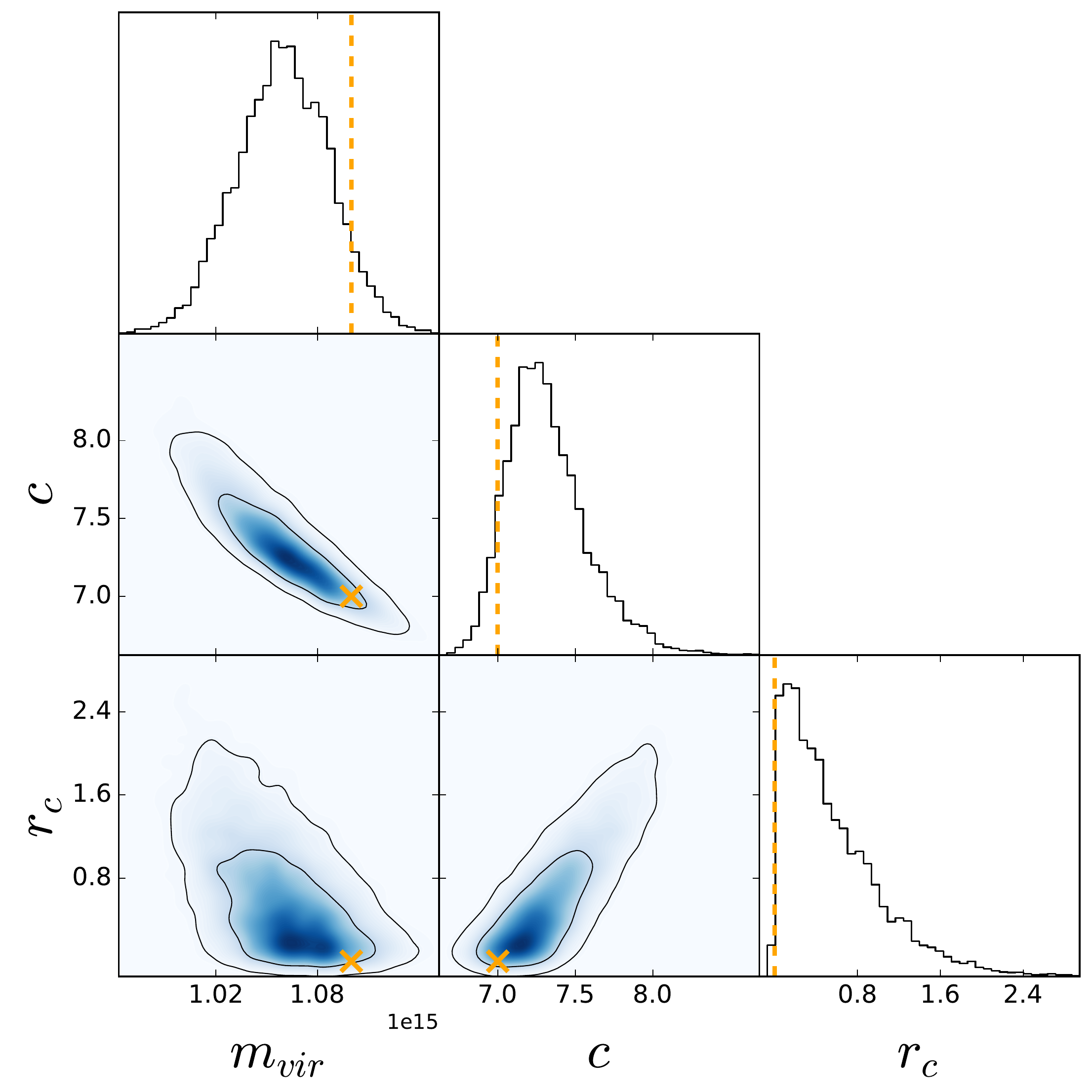}\\
    \includegraphics[width=\columnwidth]{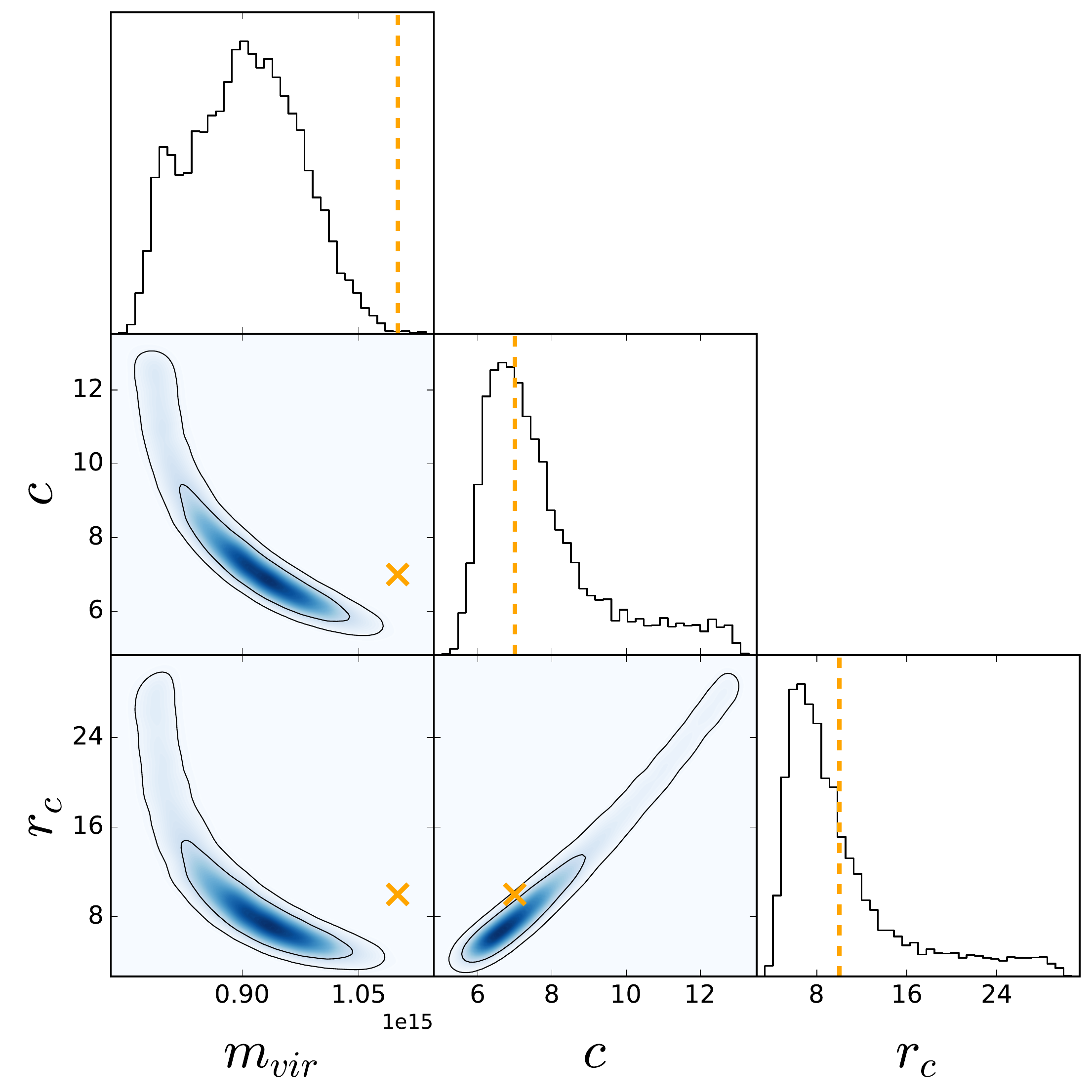}
    \includegraphics[width=\columnwidth]{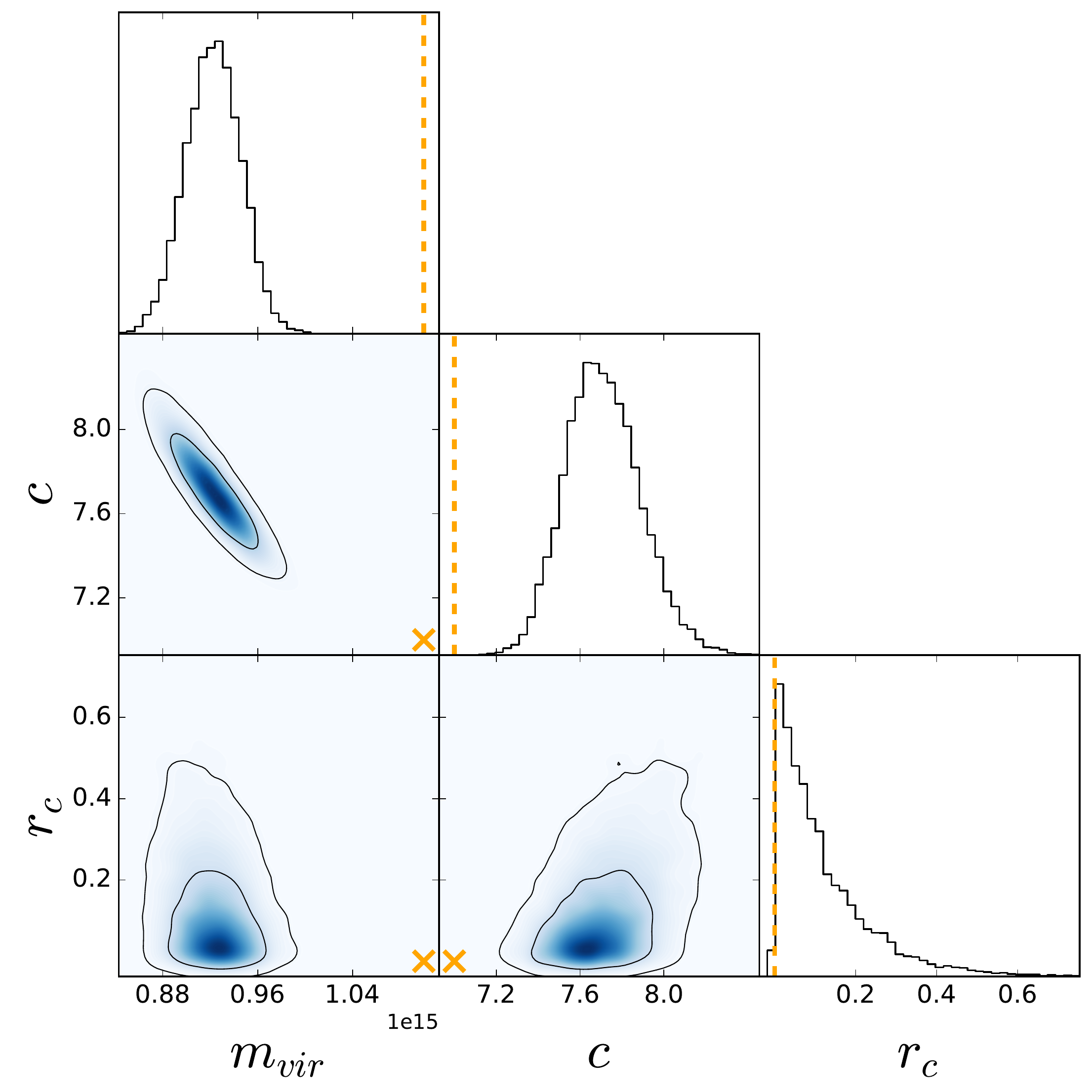}\\
    \label{fig:elliptical_approximation_params}
\end{figure*}

Turning now to the mock data sets with high ellipticity (i.e., axis ratio $q=0.5$), Figure~\ref{fig:elliptical_approximation_params} illustrates the posterior distribution results for the mass, concentration and core radius parameters. For this very elliptical halo, the models using the full elliptical density profile recover the parameters well, with the true value of all parameters located within the 1-$\sigma$ posterior contours. In contrast, the pseudo-elliptical approximation does not accurately recover the input parameters. In the cuspy case, the true value for the halo mass is outside the 2-$\sigma$ contour of the posterior. In the cored case, the true values for halo mass and concentration are both well outside the 2-$\sigma$ contours of the posteriors. As an example of how this could bias inference of core size, if weak lensing or X-ray constraints were used that constrain halo mass and concentration to be close to their proper values, this will in turn cause $r_c$ to be biased low. The lower left posterior in Figure~\ref{fig:elliptical_approximation_params} demonstrates that if the halo mass were fixed to its (correct) value of $\num{1.1e15}M_{\sun}$, the value of $r_c$ would be inferred at approximately 5\arcsec, half as large as the true value (10\arcsec). This illustrates the dangers of combining different probes to obtain core constraints if systematic errors are present in the lens model.

\begin{figure}
    \caption{Scaled surface density ($\kappa$) versus radius for a fit of a cNFW halo lens on mock data generated with a Corecusp profile. The true density profile is shown as a dotted blue line. [\textit{Top}]: Cored case, no central images. [\textit{Bottom}]: Cuspy case, no central images. Note that the accuracy of the modeled profile declines at radii far from the image locations.}
         \includegraphics[trim=10 53 10 5, clip, width=\columnwidth]{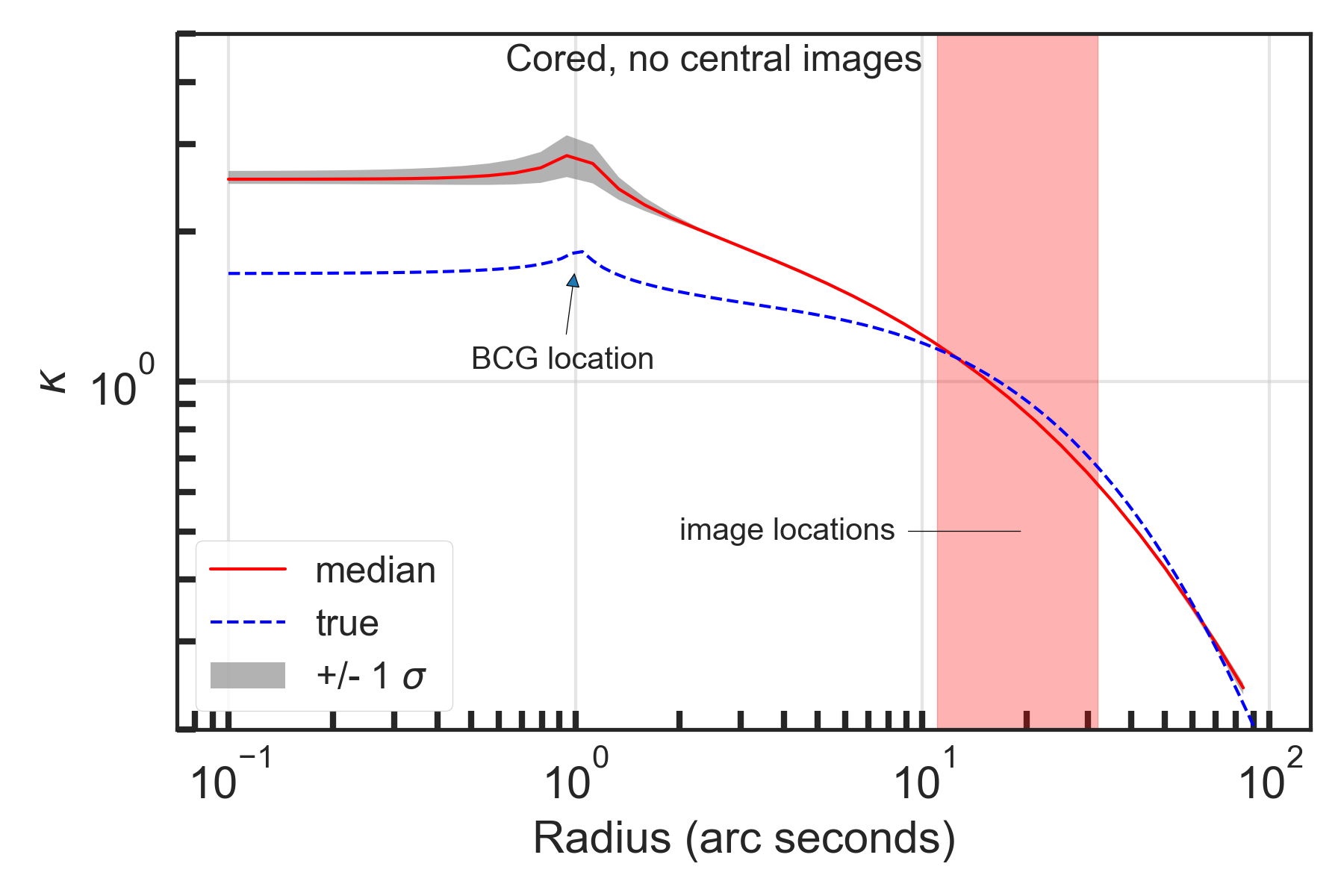}\\
    \includegraphics[trim=10 0 10 11, clip, width=\columnwidth]
    {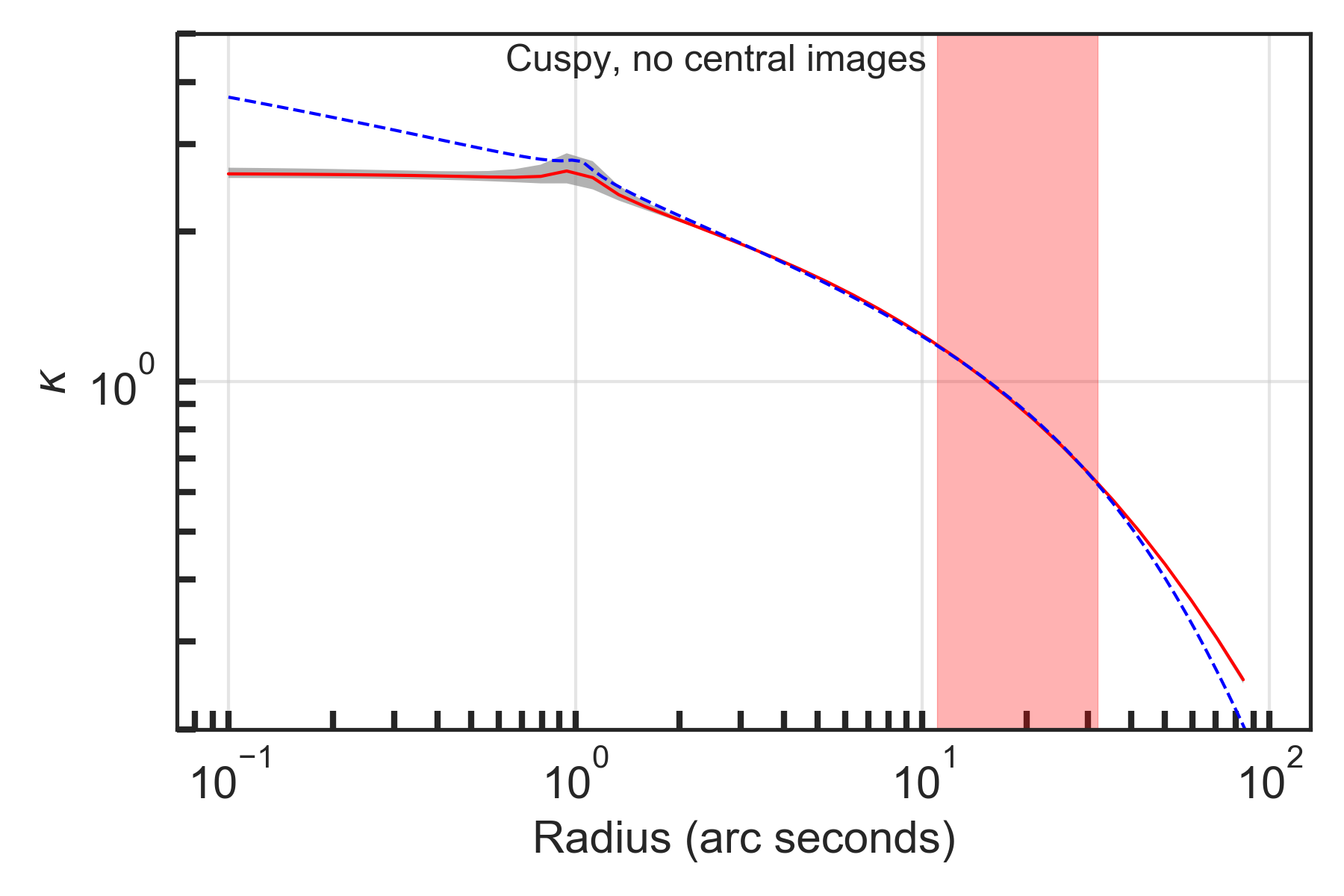}\\
    \label{fig:importance_of_shape_kappa}
\end{figure}

\subsection{Model Dark Matter Halo Radial Density Profile}
\label{Importance of Shape}
In order to explore the importance of the shape of the assumed halo profile in detecting cores, an alternative set of mock data was constructed using a Corecusp profile for the dark matter halo rather than a cNFW profile. The Corecusp profile is very similar to cNFW but has a faster transition between the inner and outer slopes. (Refer to Section~\ref{sec:Lensing Software} for definitions of the various density profiles.) The Corecusp parameters for scale radius and core radius were the same as those used in the cNFW mock data, i.e., 63\farcs7 and 10\arcsec, respectively. The slope parameters for inner and outer logarithmic slope were set to -1 and -3, respectively, matching those of a cNFW profile.

Figure~\ref{fig:importance_of_shape_kappa} shows the results of fitting a cNFW dark matter halo lens to mock data constructed with a Corecusp dark matter halo, with no central images. The fit is very good in the region of radius values of 10\arcsec to 50\arcsec, close to where the images are located. However, in attempting to fit that area as well as possible, the model predicts a cored halo for the region interior to approximately 1\arcsec, when in fact the halo is cuspy. In the case of the cored halo, the model does predict a core but overestimates the surface density in the core by approximately 40\%. Without data in the inner regions to guide it, predictions become unreliable there.

\subsection{Summary of Mock Data Results}
By modeling these mock data, it is clear that the surface density of the cluster, and thus cored or cuspy characteristics, can be well-predicted in the radial regions having image data points. Central images provide data in the inner regions (1\arcsec to 8\arcsec in our model) and thus enhance accuracy there. Using a profile shape that sufficiently approximates the true halo profile is important, as mismatches can lead to inaccurate predictions in regions devoid of image data. The use of the pseudo-elliptical approximation can lead to inaccuracies in parameter recovery exceeding 2-$\sigma$ for highly elliptical halos.

\section{Modeling of Abell 611}
\label{sec:Modeling of Abell 611}

Having tested the sensitivity of our model to varying mass profiles using strong lensing alone, we now turn to the real cluster data. We test for the presence for a core by fitting the data to two different profiles: cNFW and Corecusp.

\subsection{Abell 611 Data}
\label{Abell 611 Data}

The data for the modeling of Abell 611 were taken from sources A and B in Table A.1 of \citetalias{Donnarumma2011}. Their originally reported redshifts were 0.908 and 2.06, respectively, however subsequent analysis \citep[see][]{Newman2013a, Belli2013} indicate that the correct redshift for source A is 1.49. We have adopted that value.  Source C consists of two points at a reported redshift of 2.59, however the inclusion of this source in our model led to a large shift of the centroid of the dark matter halo, which was inconsistent with the findings from \citetalias{Donnarumma2011} and \citet{Newman2009}. This source also has the weakest photometry in the data set, with HST F606W magnitude fainter than 27.0 \citep{Richard2010}. Therefore, we decided not to include Source C in the model. Source D is a 4-image source with no confirmed spectroscopic redshift. \citetalias{Donnarumma2011} elected not to include this source in their models, and we also exclude it. We did test the inclusion of sources C and D, and found a substantial increase in the resulting $\chi^2$ of the model. We set the origin to be the coordinates of the BCG as given in Table 4 of \citetalias{Donnarumma2011} (J2000: $\ang{120.23678}, \ang{36.056572}$). The resulting data set contained 25 images in set A and 24 in set B, for a total of 49 images of 13 source points. We adopted $0\farcs2$ as the position error value, as did \citetalias{Donnarumma2011} (but see Section~\ref{The Importance of Position Errors} for a discussion of the importance of that assumption). The images are located in a range of 7 to 28\arcsec from the BCG center.

\subsection{Abell 611 Lens Model}
\label{Abell 611 Lens Model}

Following \citetalias{Donnarumma2011}, we constructed a lens model with a dark matter halo, BCG, and seven other perturbing lenses. The model parameters are described in Table~\ref{tab:Abell_611_params1}. 
Since we are using various density profiles for the dark matter halo, we need a consistent way to compare concentration and core radius, and have therefore adopted the following definitions. We define ``core radius'' as the radius at which the logarithmic slope of the density is -1, i.e., 
\begin{equation}
	\frac{d log(\rho)}{d log(r)}=-1.
\end{equation}

We define the concentration as
\begin{equation}
	\hat{c}_{200} \equiv \frac{r_{200}}{r_{-2}}
\end{equation}
where $r_{200}$ is the radius at which the density is 200 times the critical density of the universe at the redshift of the lens, and $r_{-2}$ is the radius at which the logarithmic slope of the density profile is -2. Note that for the cNFW profile, in the limit $r_c \rightarrow 0$ where the profile reduces to NFW, we have $r_{-2} = r_s$.

The Abell 611 system was analyzed using the MultiNest sampling algorithm \citep{Feroz2009} with 4,000 live points. There were 14, 15 and 13 free parameters for the cored NFW, Corecusp and NFW profiles, respectively. Considering the 49 image points being generated from 13 sources, we then have 57 to 59 degrees of freedom depending on the lens model. We tested both source plane and image plane chi-square evaluations, and found very similar results for each method. Source plane chi-square evaluations were used for all runs, with additional image plane evaluations made as needed to verify the correct reproduction of multiple images.

\subsubsection{Cluster Halo and BCG Models}

We studied three mass profiles for the cluster halo: cNFW, Corecusp and NFW. Uniform priors were used on all free halo parameters except the core scale parameter $r_{c, kpc}$, for which a log prior was used. The BCG was modeled as a dPIE profile with mass parameter b left free, and other parameters fixed at the values given by \citet{Newman2013a}. Table \ref{tab:Abell_611_params1} summarizes the parameter values and ranges.

\begin{table}
 \caption{The parameters and prior ranges for the dark matter halo, BCG and seven perturbing galaxies in the Abell 611 Lens Model. All priors are uniform over their range except for the parameter $r_{c, kpc}$, for which a log prior was used.}
	\centering
		\begin{tabular}{llcc} % columns, alignment for each
		\hline
		Parameter & Description & Units & Prior Range \\
		\hline
        \multicolumn{3}{l}{\textit{DM Halo (cNFW, Corecusp and NFW)}}\\
		$M_{200}$ & halo mass & $10^{14}M_{\sun}$  & 3 - 20 \\
        $c_{200} $ & concentration &  -   & 1 - 40 \\
        $r_{c, kpc}$ (*) & core scale & kpc & 0.001 - 500 \\
        $\gamma$ (**) & inner log slope & - & 0 - 2.99 \\
        q & axis ratio & -  & 0.5 - 0.95 \\
        $\theta$ & orientation & deg.  & 120 - 160 \\
        x-center & x coord. of center &\arcsec & -5 to 5\\
        y-center & y coord. of center &\arcsec & -5 to 5 \\
        \\
        \multicolumn{2}{l}{\textit{BCG (dPIE profile)}}\\
        b & mass parameter &\arcsec & 0.5 - 10.0 \\
        a & scale radius &\arcsec & fixed: 10.7 \\
        s & core radius &\arcsec& fixed: 0.277 \\
        q & axis ratio & - & fixed: 0.73 \\
        $\theta$ & orientation & deg. & fixed: 132.3 \\
        x-center & x coord. of center &\arcsec & fixed: 0.0 \\
        y-center & y coord. of center &\arcsec & fixed: 0.0\\ 
               \\
        \multicolumn{2}{l}{\textit{Cluster Members (dPIE profile)}}\\
        b & mass parameter &\arcsec & 0.05 - 10.0 \\
        a & scale radius &\arcsec & 0.05 - 10.0 \\
        s & core radius &\arcsec& fixed ($\dagger$) \\
        q & axis ratio & - & fixed ($\dagger$) \\
        $\theta$ & orientation & deg. & fixed ($\dagger$) \\
        x-center & x coord. of center &\arcsec & fixed ($\dagger$)\\
        y-center & y coord. of center &\arcsec & fixed ($\dagger$)\\ 
		\hline
        \end{tabular}
      		\begin{footnotesize}
            \begin{flushleft}
             * cNFW and Corecusp only\\
             ** Corecusp only\\
      		 $\dagger$ See Table~\ref{fig:Abell_611_params2} for these values.\\
             \end{flushleft}
            \end{footnotesize}
	\label{tab:Abell_611_params1}
\end{table}
\subsubsection{Cluster Member Models}
The seven perturbing lens elements were modeled with dPIE profiles, allowing for separate specification of their mass, core size, cutoff radius, axis ratio, orientation angle and centroid. Perturbers 1 and 2 are quite close to image groups B.4 and B.5, allowing a stronger constraint on their Einstein radii. As such, the mass and cutoff radius parameters for those perturbers were left free. To avoid a proliferation of parameters, the mass and cutoff parameters for perturbers 3 through 7 were anchored together, allowing two parameters to specify the mass and scale for that group.

The ``b" parameter of the dPIE lens is proportional to the lens mass and varies as the square of velocity dispersion (see Equation \ref{eq: b}). \citet{Faber1976a} show that velocity dispersion scales as $L^{1/4}$, so the relevant scaling relation is 
\begin{equation}
	b^\prime = b \bigg(\frac{L^\prime}{L}\bigg)^\frac{1}{2}
\end{equation}
The mass parameters for perturbers 3 through 7 are anchored together according to this relation. The luminosities are shown in Table~\ref{fig:Abell_611_params2}.

Similarly, the cutoff radii and core radii were scaled using 
\begin{equation}
	r_{cut}^\prime = r_{cut} \bigg(\frac{L^\prime}{L}\bigg)^\frac{1}{2}
\end{equation}
and
\begin{equation}
	r_{c}^\prime = r_{c} \bigg(\frac{L^\prime}{L}\bigg)^\frac{1}{2}
\end{equation}
The exponent $1/2$ in the two equations above correspond to a constant mass-to-light ratio among perturbers 3 through 7. The cutoff radius normalization for those perturbers was a free parameter. As the core radii are difficult to constrain without visible images near the core of the cluster member, they were fixed according to a normalized core radius of 0\farcs035 for an ST magnitude of 18.0, matching the assumption of \citetalias{Donnarumma2011}. For the the centroid locations, axis ratios, and orientations of the perturbers, \citetalias{Donnarumma2011} used GALFIT to determine those values, and we adopt them. They are shown in Appendix Table~\ref{fig:Abell_611_params2}.

\begin{figure*}
    \caption{Image plane representation (left, with critical curves shown for z=1.49), and source plane representation (right, with caustic curves shown for z=1.49) of the best fit result for the Abell 611 cNFW model. The data points are shown in red, the modeled images in cyan, and show purple where they overlap.}
          \includegraphics[width=0.49\textwidth]{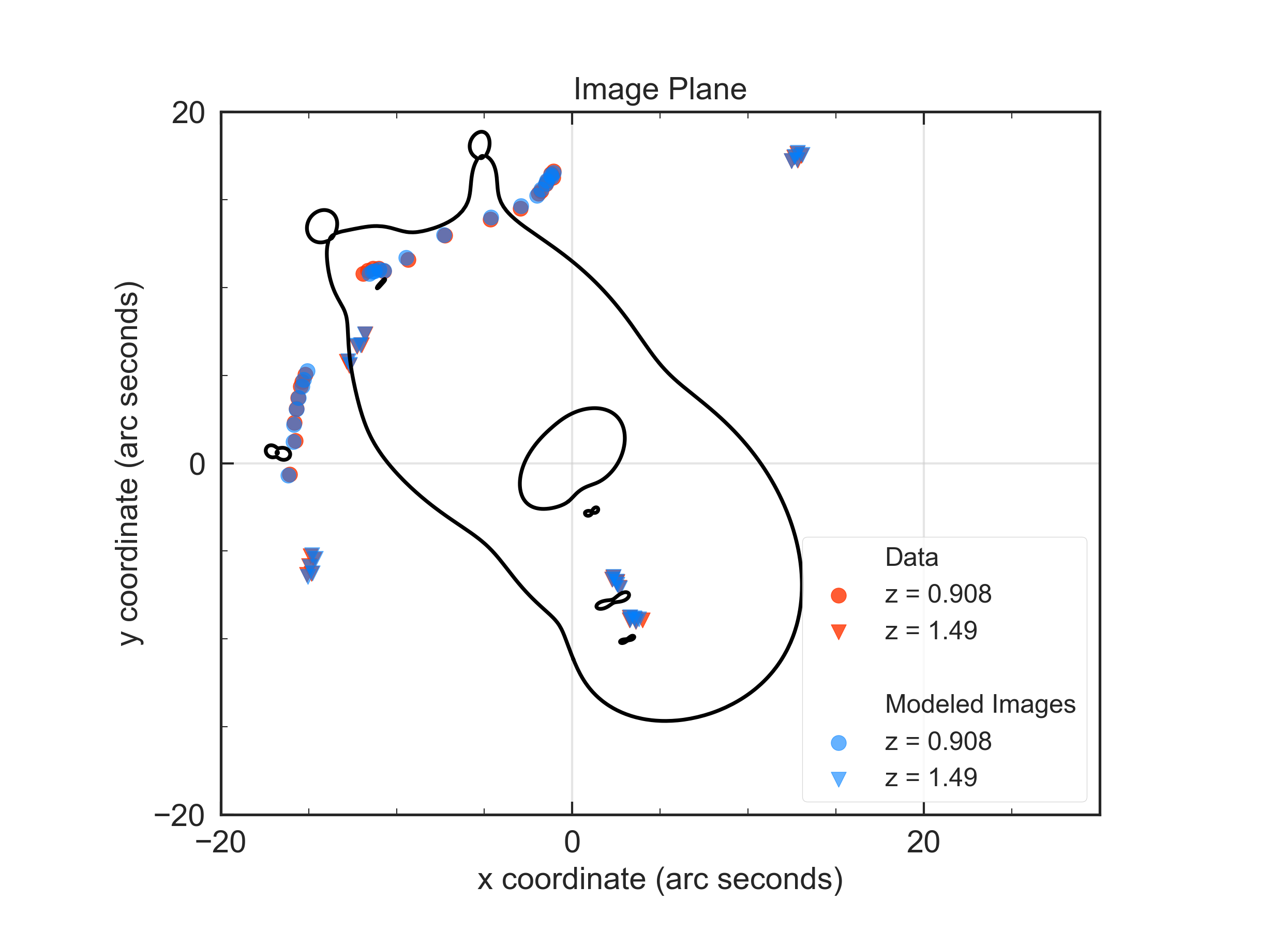}
      \hfill
      \includegraphics[width=0.49\textwidth]{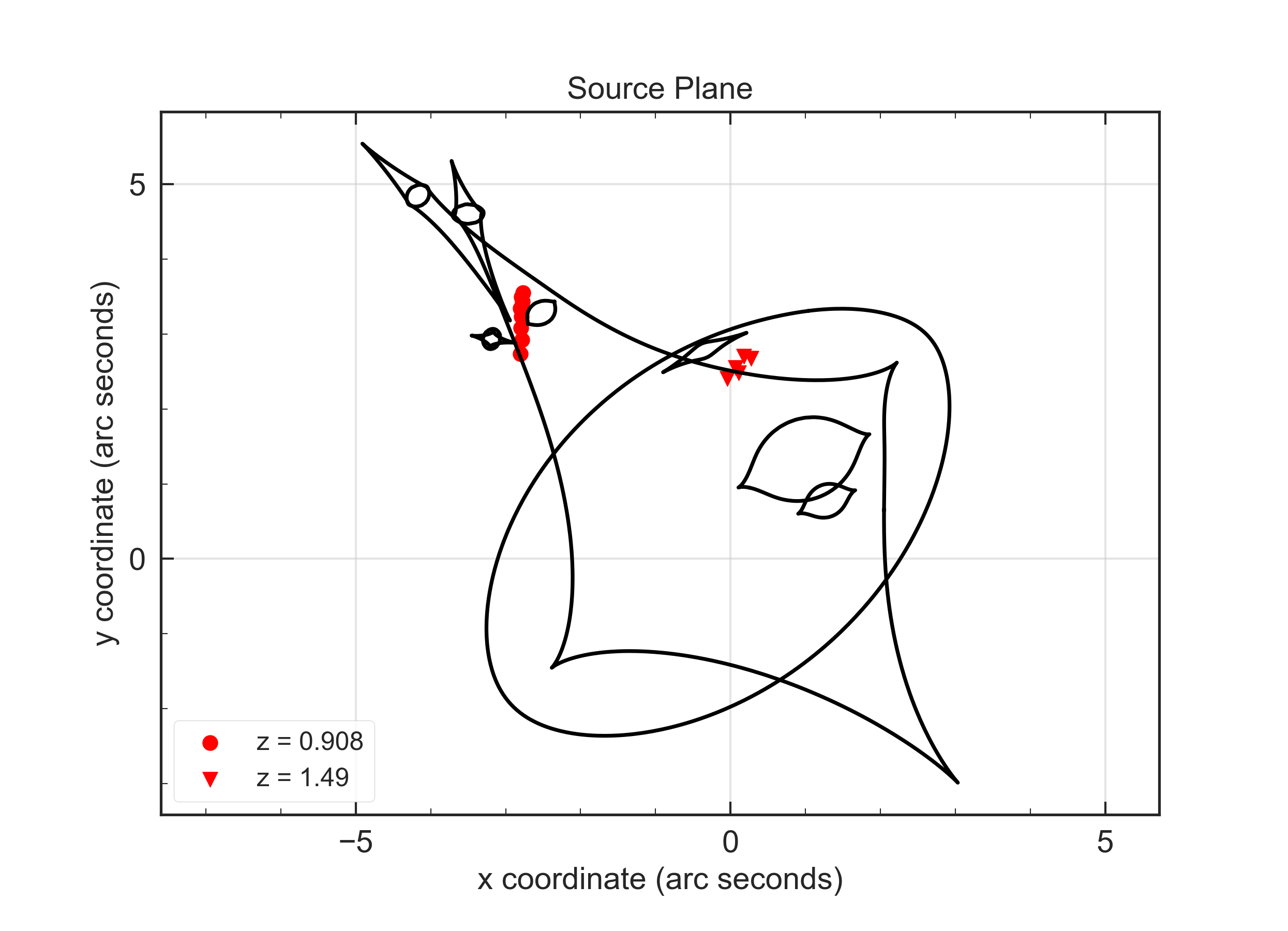}\\
    \label{fig:Abell_611_fit}
\end{figure*}

\subsection{Abell 611 Results}
\label{Abell 611 Results}

The resulting fits were very good for all the profiles modeled. The reduced $\chi^2$ for the fits range from 0.28 and 0.30. The cNFW is our baseline model and is marginally favored by the Bayesian evidence (ln(Z) = -44.8) over the Corecusp model (ln(Z) = -46.6). The resulting best-fit images for the cored NFW profile model are shown in Figure~\ref{fig:Abell_611_fit}, and the key best-fit parameters for all models are shown in Table~\ref{tab:Abell_611_results}. 

\begin{table*}
	\centering
    \caption{Key median posterior parameter values for the Abell 611 models. The bounds of the 68\% confidence interval are also shown. The cNFW and Corecusp models exhibit bimodal solutions for some parameters, allowing either a near-zero core or a large core of 14\arcsec to 16\arcsec.}
         \renewcommand{\arraystretch}{1.6}
		\begin{tabular}{llccccc} % columns, alignment for each
         Halo profile  &       & \multicolumn{2}{c}{cNFW} & \multicolumn{2}{c}{Corecusp} & NFW \\
         
         \hline
		 Mode & & Sm. Core & Lg. Core & Sm. Core & Lg. Core & (no core) \\ 
		\hline
        \underline{Bimodal Parameters} & \underline{Units} \\
		$M_{200}$ & $10^{14}M_{\sun}$  & $12.7_{-1.1}^{+1.4}$ & $8.4_{-0.9}^{+1.7}$ & $7.2_{-0.6}^{+0.7}$ & $5.4_{-0.6}^{+0.8}$ & $12.8_{-1.3}^{+1.6}$ \\
        $\hat{c}_{200} $ &  -   & $4.1_{-0.4}^{+0.3}$ & $6.1_{-0.9}^{+0.6}$ & $6.3_{-0.4}^{+0.4}$ & $7.6_{-0.6}^{+0.6}$ & $4.1_{-0.4}^{+0.4}$ \\
        $r_{core}$ &\arcsec & $0.3_{-0.2}^{+1.2}$ & $13.3_{-2.0}^{+1.3}$ & $0.01_{-0.01}^{+0.6}$ & $15.6_{-2.6}^{+1.5}$ & (0 by defn.)  \\
        $M_{BCG}$ & $10^{12}M_{\sun}$ & $4.9_{-1.0}^{+1.2}$ & $7.2_{-1.0}^{+1.2}$ & $4.7_{-0.9}^{+1.0}$ & $6.8_{-1.0}^{+1.2}$ & $4.9_{-1.1}^{+1.3}$ \\
        $\kappa_{tot}(5\arcsec)$ & -  & $1.32_{-0.01}^{0.01}$ & $1.30_{-0.01}^{+0.01}$ & $1.32_{-0.01}^{+0.01}$ & $1.29_{-0.01}^{+0.01}$ & $1.32_{-0.01}^{+0.01}$ \\
        $\kappa_{DM}(5\arcsec)$ & -  & $1.12_{-0.04}^{+0.04}$ & $1.02_{-0.04}^{+0.04}$ & $1.13_{-0.03}^{+0.03}$ & $1.03_{-0.04}^{+0.04}$ & $1.12_{-0.04}^{+0.04}$ \\
        $\kappa_{DM}(20\arcsec)$ & -  & $0.60_{-0.01}^{+0.01}$ & $0.59_{-0.01}^{+0.01}$ & $0.60_{-0.01}^{+0.01}$ & $0.59_{-0.01}^{+0.01}$ & $0.60_{-0.01}^{+0.01}$ \\
        $\gamma$ (Corecusp only) & - &  &  & $1.04_{-0.03}^{+0.05}$ & $0.71_{-0.19}^{+0.16}$ &  \\
\\
        \underline{Unimodal Parameters} & \underline{Units} \\
    	position angle & degrees & \multicolumn{2}{c}{$133.3_{-0.2}^{+0.2}$} & \multicolumn{2}{c}{$133.3_{-0.2}^{+0.3}$}  & $133.3_{-0.2}^{+0.3}$ \\
        axis ratio & - & \multicolumn{2}{c}{$0.67_{-0.01}^{+0.01}$} & \multicolumn{2}{c}{$0.67_{-0.01}^{+0.01}$} & $0.67_{-0.01}^{+0.01}$ \\
        x-center &\arcsec &  \multicolumn{2}{c}{$-0.2_{-0.3}^{+0.3}$} & \multicolumn{2}{c}{$-0.3_{-0.3}^{+0.3}$} &  $-0.2_{-0.2}^{+0.2}$ \\
        y-center &\arcsec & \multicolumn{2}{c}{$0.7_{-0.3}^{+0.3}$} & \multicolumn{2}{c}{$1.0_{-0.3}^{+0.3}$} & $0.7_{-0.2}^{+0.3}$ \\ 
		\hline    
        \end{tabular}
    \renewcommand{\arraystretch}{1} 
	\label{tab:Abell_611_results}
\end{table*}

\begin{figure}
\caption{Scaled surface density ($\kappa$) versus radius for Abell 611. cNFW and Corecusp models are shown, each subdivided into large-core and small-core solutions.}
\includegraphics[width=\columnwidth]{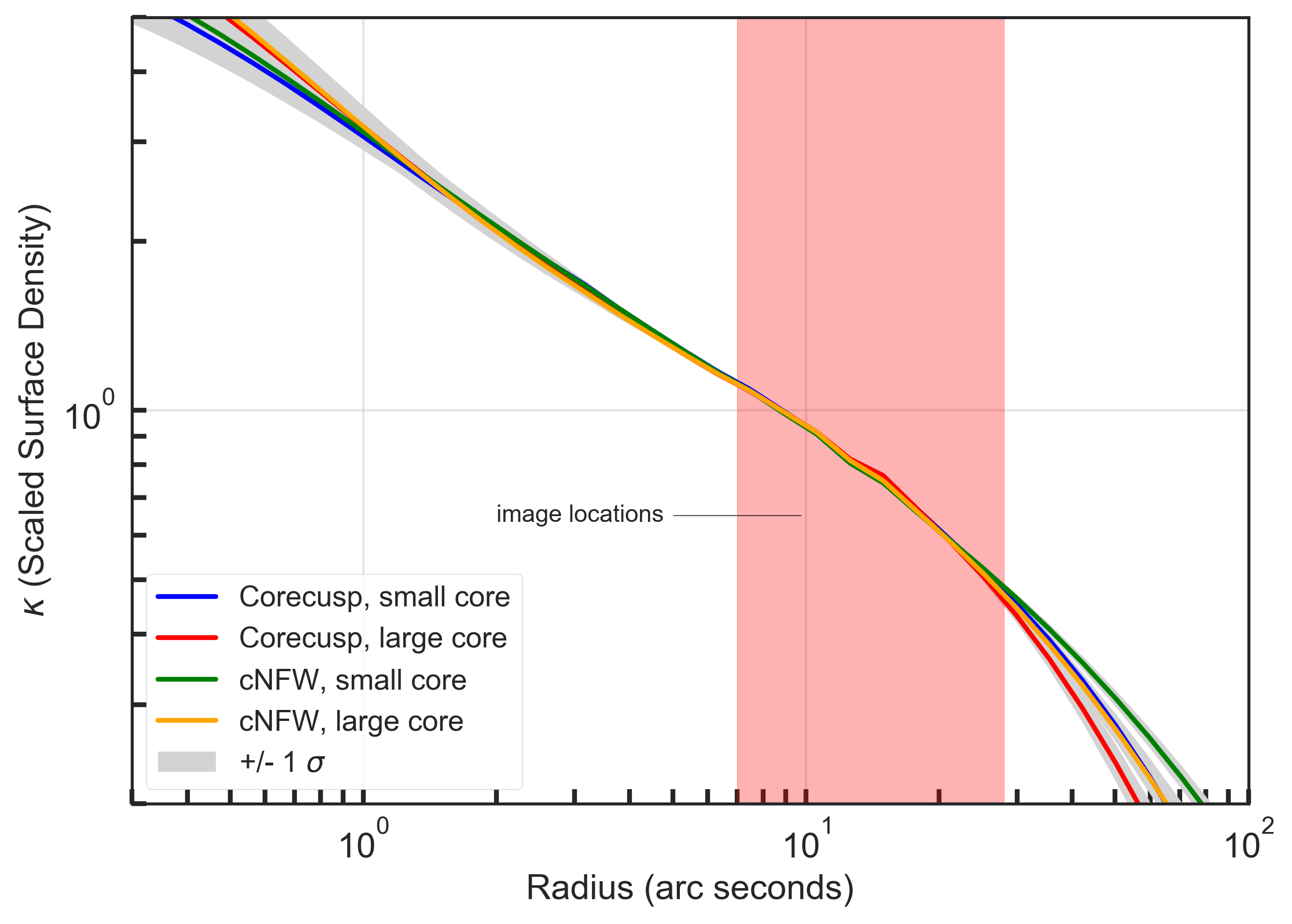}
\label{fig:A611_kappa}
\end{figure}

The posterior distributions for the cored NFW and Corecusp models both exhibit a bimodal solution for the lens model parameters. There is a clear ``small-core" mode, with a core radius < 1\arcsec, and a ``large-core" mode, with core radius $\sim15$\arcsec. Two-dimensional posterior distribution plots for selected parameters for the cNFW and Corecusp cases are included in Appendix Figure~\ref{fig:A611_cNFW_triangle} and Figure~\ref{fig:A611_cc_triangle}, respectively. The $\chi^2$ for the best fit points of each of the two modes are very similar: 16.5 for the small-core cNFW mode and 17.1 for the large-core cNFW mode. The small-core mode is associated with a higher halo mass and a lower BCG mass, while the large-core mode is reversed in that regard. 

There is clearly a degeneracy between the halo core and BCG mass, as the halo and BCG are nearly co-centered (approximately 1\arcsec apart), and it is the combination of their masses that determines the surface density and hence deflection angles. In evaluating these two modes, we can ask whether the resulting BCG mass is consistent with prior constraints on early-type galaxies. The luminosity of the BCG was found to be $\num{5.47e11}M_{\sun}$ in V-band \citep{Newman2013a}, which, given the median BCG masses in Table~\ref{tab:Abell_611_results} would imply best-fit stellar mass-to-light ratios of 9 and 13 for the small-core and large-core modes, respectively. At the low-mass end, we infer the small-core BCG mass $\gtrsim \num{3.0e12}M_{\sun}$ at 95\% CL, equating to a minimum stellar mass-to-light ratio of 5.5; for comparison, the large-core mode requires a BCG mass $\gtrsim \num{5.6e12}M_{\sun}$ at 95\% CL, equating to a stellar mass-to-light ratio of 10.3. Such high mass-to-light ratios imply that both solutions require a steep stellar initial mass function (IMF). However, as we will show in Section \ref{sec:imf}, the IMF slope required by the small-core mode is consistent with recent constraints from high-mass early-type galaxies, whereas the large-core mode is inconsistent with these constraints.

As might be expected, the NFW halo model produces nearly identical posteriors to the small-core mode cNFW solution, albeit with a slightly higher $\chi^2$ (17.0 versus 16.3 for cNFW).
The concentration $\hat{c}_{200}$ is 4.1 for the small core solution, consistent with previously studied mass-concentration relations \citep{Neto2007}, although it should be noted that those relations were created for systems with NFW profiles, and may not be easily compared to other forms of profiles that have cores.  Interestingly, the small-core mode of the Corecusp model prefers an inner slope of 1.05, very similar to an NFW inner slope. However, the Corecusp solution is more concentrated ($\hat{c}_{200}=6.2$) and has a much smaller halo mass, as can be seen in Table~\ref{tab:Abell_611_results}.

The resulting posterior distributions for many parameters are similar between the cored NFW profile and the Corecusp profile, and are unimodal. These include orientation angle ($\theta$), centroid location ($x_c$, $y_c$), axis ratio (q) and scaled surface density ($\kappa$). In particular, $\kappa_{tot}$ at 5\arcsec is very well constrained and is remarkably consistent between the models, varying between 1.28 and 1.32. A plot of $\kappa$ versus radius is shown in Figure~\ref{fig:A611_kappa}, with cNFW and Corecusp posteriors separated into their large-core and small-core components. Their median values are in close agreement in the range of radii between 1\arcsec and 30\arcsec.

\subsubsection{Implications for the Stellar Initial Mass Function of the BCG}\label{sec:imf}

\begin{figure}
    \caption{Mass-to-light mismatch parameter $\alpha_{Salp}$ versus IMF slope $\Gamma$, where $\alpha_{salp}$ is defined as the ratio of $M_*/L_V$ produced by this IMF over the value expected from SPS models (found in Newman et al. 2013) using a Salpeter IMF. Constraints from galaxy surveys in Cappellari+ 2013, Conroy+ 2017 and Leier+ 2016 are overlaid. The right panel shows the posterior probability densities in $\alpha_{salp}$ for the small-core and large-core solutions.}
    \includegraphics[width=0.49\textwidth]{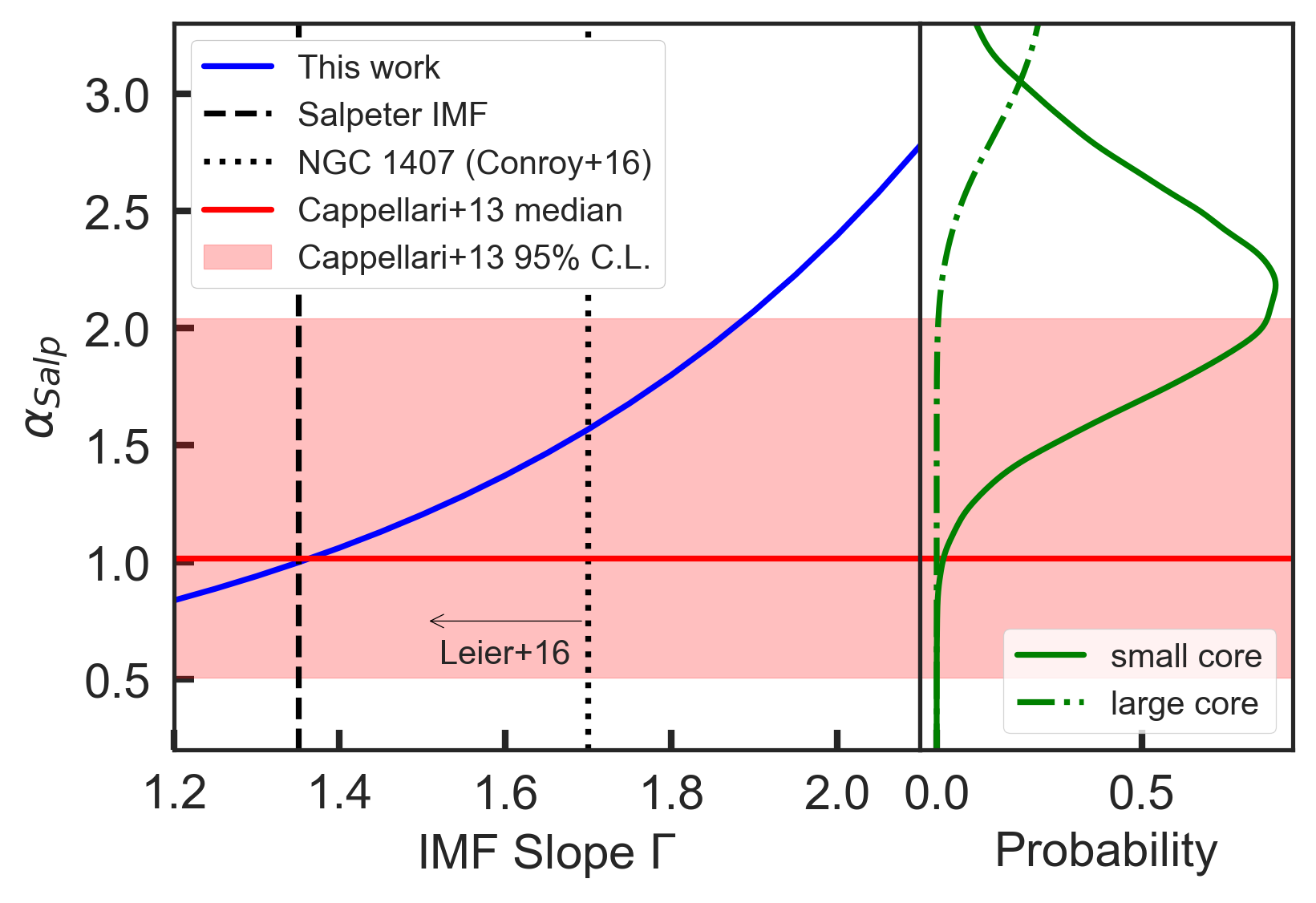}
    \label{fig:ML_vs_IMF}
\end{figure}

The high stellar mass-to-light ratio we have inferred for the BCG would suggest a fairly steep stellar initial mass function. Given the high central dispersion of the BCG, this is not suprising: many authors in recent years have inferred a bottom-heavy IMF in early-type galaxies, using either spectral lines \citep{barbera2013,cappellari2013,conroy2017,lyubenova2016,conroy2012} or strong lensing \citep{leier2016,Newman2013a}. In \cite{barbera2013} and \cite{cappellari2013}, spectra from a large sample of early-type galaxies (SPIDER and ATLAS-3D, respectively) were analyzed, revealing a trend in the IMF log-slope: galaxies with low central dispersions show a shallow slope consistent with a Chabrier/Kroupa IMF, whereas galaxies with higher dispersions show a steeper slope, comparable to or even steeper than that of a Salpeter IMF.

This begs the question, are either our small- or large-core solutions compatible with constraints on the IMF in early-type galaxies? These solutions require stellar mass-to-light ratios of at least $M_*/L_V \gtrsim 5.5$ and $\gtrsim 10.3$, respectively (at 95\% CL). Since $M_*/L_V$ depends on other factors besides the IMF (e.g. metallicity, stellar ages), one way to compare IMF constraints is to define the ``IMF mismatch'' parameter $\alpha_{salp} = (M_*/L_V) / (M_*/L_V)_{salp}$, where $(M_*/L_V)_{salp}$ is the mass-to-light ratio generated by a Salpeter IMF. Positive $\alpha_{salp}$ values then would imply an IMF that is more bottom-heavy compared to Salpeter. In \cite{Newman2013a}, the stellar mass-to-light ratio $M_*/L_V$ for Abell 611 was estimated using the BCG colors and a stellar population synthesis model. Under the assumption of a Salpeter IMF, they infer $(M_*/L_V)_{salp} = 3.98$. Using this, our small- and large- core solutions require $\alpha_{salp}$ values of at least 1.38 and 2.59, respectively, at 95\% CL.

In \cite{cappellari2013} a trend line is fit to $\log\alpha_{salp}$ as a function of dispersion (see their Figure 13), with corresponding lines for 2.6$\sigma$ scatter. Using the fact that the luminosity-weighted dispersion of the BCG within its half-light radius is 306 km/s, we estimate the median value for $\alpha_{salp} \approx 1.0$ with the $\pm2.6\sigma$ bounds at 0.5 and 2.0. \cite{lyubenova2016} find a similar range using galaxies in the CALIFA survey, for which $\alpha_{salp}$ lies in the approximate range 0.6-1.5, while \cite{leier2016} find a similar range 0.5-1.5 in the SLACS lens sample. Given these constraints, it is evident that our small-core solution ($\alpha_{salp} \gtrsim 1.4$) is compatible with current constraints, whereas the $\alpha_{salp} \gtrsim 2.6$ required by the large-core solution lies beyond the upper bound for all the surveys mentioned here.

Next, we go further and estimate the constraint on the slope of the IMF of the BCG in Abell 611 from our lensing analysis. We will model the IMF using a double power-law model, $\xi(M) \propto M^{1-\Gamma}$ where $\Gamma = 1.35$ (Salpeter) for $M > M_\odot$ while $\Gamma$ for $M < M_\odot$ will be freely varied. This is identical to one of the models used in \cite{leier2016} and one of the parametric models used in \cite{conroy2017}.
 To relate the mass-to-light ratio to the IMF slope $\Gamma$, we have

\begin{equation}
\frac{M_*}{L_V} = \frac{\int_{M_{low}}^{M_{high}}M^{2-\Gamma}}{\int_{M_{low}}^{M_{high}}L_V(M) M^{1-\Gamma}}
\label{m_over_l_formula}
\end{equation}
where $L_V(M)$ is the V-band stellar luminosity-mass relation. For the lower mass cutoff we adopt the usual convention $M_{low} = 0.1 M_\odot$, and the high mass cutoff $M_{high}$ will be determined by the particular isochrone used. For the mass-luminosity relation we use the Padova isochrones \citep{girardi2004} assuming metallicity $Z = Z_\odot$ (the same choice was adopted in \cite{Newman2013a}), and consider a few different stellar ages. To account for the fact that the stellar ages inferred for Abell 611 may differ from our choices here (along with possible slight differences in the SPS model used), we will write $L_V(M) = \lambda L_{V,0}(M)$ where $L_{V,0}(M)$ is generated from the isochrone, and $\lambda$ is a correction factor which we expect to be close to 1 if the correct median stellar age is assumed. For a given assumed stellar age, we perform the integration in Eq.~\ref{m_over_l_formula} by interpolating over a table of values in $L_{V,0}(M)$ generated from the isochrone, then solve for $\lambda$ using the \cite{Newman2013a} values ($M_*/L_V = 3.98$, $\Gamma = 1.35$). With $\lambda$ in hand, we can then use Eq.~\ref{m_over_l_formula} to plot the IMF slope $\Gamma$ needed to produce a given stellar mass-to-light ratio. In practice, we find that the results are nearly identical regardless of stellar age (we tried ages in the range of 6-10 Gyr), since the luminosity is sensitive to the IMF slope at high stellar mass which is still Salpeter in our model; since $\lambda \approx 1$ for 8 Gyr, we adopt this stellar age in the following.

In Figure \ref{fig:ML_vs_IMF} we plot $\alpha_{salp}$ as a function of IMF slope $\Gamma$, while on the right side is plotted the posteriors of the small- and large-core solutions in $\alpha_{salp}$. Note that the curve equals 1 at $\Gamma = 1.35$, since the above procedure enforces consistency with the \cite{Newman2013a} results. From this figure we see that an IMF slope $\Gamma \approx 1.5$ is required to produce a $\alpha_{salp} \approx 1.4$, which is close to the 95\% CL lower limit required by our small-core solution. By contrast, a slope $\Gamma \gtrsim 2$ is required to be consistent with the large-core solution. Among the SLACS lenses, \cite{leier2016} found that for the double power law IMF, a slope of 1.7 implies a dark matter fraction of zero, and hence steeper slopes are ruled out. Indeed, \cite{chabrier2014} argue that in the most extreme starburst conditions, the IMF ``saturates'' at a slope $\Gamma \approx 1.7$. Recently \cite{conroy2017} investigated a galaxy with a similar dispersion to ours (NGC 1407) and found $\Gamma = 1.7$, possibly reaching the saturation limit. While the small-core solution is consistent with these constraints, the large-core solution clearly is not, painting a consistent picture with the above constraints on $\alpha_{salp}$.

Our inference of a stellar IMF slope $\Gamma \gtrsim 1.5$ carries some important caveats. First, any inference about the IMF slope depends on the form of the IMF used. A popular variant is the ``bimodal'' IMF \citep{vazdekis1996}, favored in several recent studies \citep{lyubenova2016,leier2016,barbera2016}, which uses a variable slope $\Gamma_b$ at $M > M_\odot$ whereas the slope tapers to zero at low masses. As an additional check, we repeated the above analysis using the bimodal IMF and found that a slope $\Gamma_b \gtrsim 3$ is required for the small-core mode. This is near the upper limit of the ranges observed in surveys (\citealt{lyubenova2016} find a maximum $\Gamma_b \approx 3.1$, while \citealt{barbera2013} infer $\Gamma_b \sim$ 3.0 for $\sim$ 300 km/s dispersions); again, the large-core mode requires a much higher slope and hence is likely ruled out.

Another caveat is that the IMF slope may vary with radius, as recent studies have suggested \citep{barbera2016,zieleniewski2017,vandokkum2017}. This is important because the presence of the BCG in the total projected density is only distinct out to $\sim 0.3$ times the effective radius, as can be seen from the size of the ``bump'' in Figure \ref{fig:A611_kappa} (note that the effective radius is $\sim$ 10 arcsec). Thus, we may only be sensitive to the stellar mass in the inner regions, where the mass-to-light ratio is high. If the IMF indeed becomes shallower further out, then the \emph{total} $M_*/L_V$ may potentially be lower than we infer from the strong lens modeling. It would also imply that the stellar mass profile becomes steeper than the light profile at larger radii, which could be an important systematic when inferring stellar masses from lensing. Allowing for a possible steepening of the stellar mass profile relative to the light profile when doing the lens modeling (to account for this systematic) is left to future work.

\subsubsection{Performance of the Pseudo-Elliptical Model}

The pseudo-elliptical cNFW model yielded a best-fit $\chi^2$ that was slightly higher than the corresponding elliptical cNFW fit, and the resulting posteriors were generally similar for most parameters. Interestingly, the median posterior value for the BCG mass was approximately 50\% higher when using the pseudo-elliptical approximation. This is remarkable given that the ellipticity is not extreme: the inferred ellipticity of the potential contours is $\epsilon \approx 0.19$ (where $\epsilon$ is defined the same as in \citealt{Golse2002}), which is low enough that it might appear ``safe'' to use the pseudo-elliptical approximation. By contrast, the cNFW fit inferred an axis ratio $q \approx 0.67$ for the density contours, markedly lower than one might have naively expected from the pseudo-elliptical fit. (We note that in our mock data runs, there did not appear to be systematic difference in the BCG mass when using the pseudo-elliptical approximation, even in the case of high ellipticities.)

The inferred BCG mass using the pseudo-elliptical model makes the best-fit stellar mass-to-light ratio even higher ($\approx$ 13 for the small-core mode), making it much harder to reconcile with IMF constraints for early-type galaxies. We conclude that the pseudo-elliptical model can bias the results significantly even if the inferred ellipticity is not extreme, and hence modeling lenses with true elliptical density contours is strongly preferred.

\subsubsection{Consideration of Stellar Kinematic Data}

\citet{Newman2013a,Newman2013b} used long-slit spectroscopic observations of velocity dispersion in the BCG and spherical Jeans equation analysis to find a $\chi^2$ for those stellar kinematic data, which is then incorporated into their overall fit. The attraction to this approach is that it incorporates constraints from the inner region of the cluster, where there are no strong lensing images due to the bright BCG image. They assume that the BCG is centered at the same location as the dark matter halo and that the system is spherical. For their fiducial case they assume an isotropic system, i.e., $\beta_{aniso}=1-(\sigma^2_{\theta}/\sigma^2_r)=0$, but they also ran models for $\beta_{aniso}$ values between -0.2 and +0.2, with constant values of $\beta$ in all cases. We adopt their dispersion observations and error values. However, we excluded the innermost point from the analysis, as that point is subject to systematic error from slit and seeing effects that are greater than the observational error (A. B. Newman, personal communication, December 7, 2018). We then apply the spherical Jeans analysis.

As a starting point, we used a cNFW model similar to our baseline but adopt a fixed BCG mass of $\num{1.5e12}M_{sun}$, which is similar to the mass found in \citet{Newman2013a} and \citetalias{Donnarumma2011}. We then produced a ``large core" chain, with $r_{core}>10$\arcsec, and a ``small core" chain, with $r_{core}<3$\arcsec. We analyzed an isotropic case with $\beta=0$ as well as mildly radially and tangentially biased cases with $\beta=\pm0.2$.

Following \citet{Capellari2008}, the velocity dispersion over the line of sight can be found from
\begin{equation}
\sigma_{BCG,LOS}^2(R)=\frac{2G}{\Sigma_{BCG}(R)} \int_R^{\infty } \frac{\mathcal{F}(r,R,\beta)  \rho_{BCG}(r)   M(r)}{r^{2-2 \beta }} dr,
\end{equation}
where $\Sigma_{BCG}$ is the BCG surface density (derived in our case from the 3D dPIE profile), $\rho_{BCG}$ is the dPIE density profile, $M(r)$ is the mass of all components generating the potential and $\mathcal{F}(r,R,\beta)$ is an analytic function derived in \citet{Capellari2008}.

\begin{figure}
    \caption{Velocity dispersion of the BCG, and corresponding fit to the observations of \citet{Newman2013a}, assuming $\beta=0$.}
         \includegraphics[width=\columnwidth]{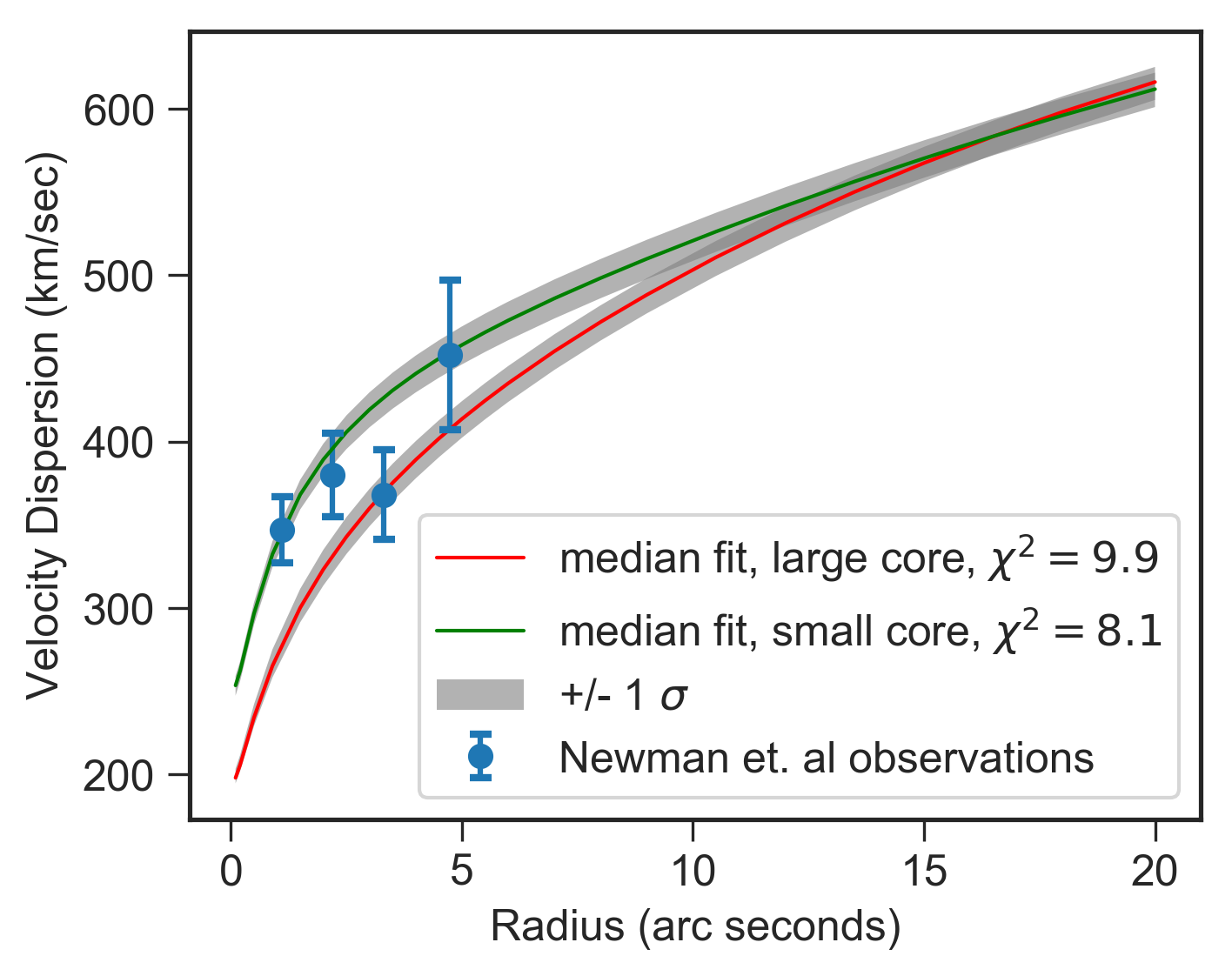}\\
    \label{fig:dispersion}
\end{figure}

The velocity dispersions assuming $\beta=0$ at all radii is shown in Figure~\ref{fig:dispersion}. The small core and large core cases both provide plausible fits to the stellar kinematic data. As $\beta$ is varied over a modest range of -0.2 to +0.2, the fits change from favoring a small core to favoring a large core. These represent a constant value of $\beta$ at all radii; if $\beta$ were allowed to vary with radius, even within these modest bounds, a wide variety of solutions could be accommodated. In \citet{Schaller2015}, they examined six simulated clusters similar in size and character to Abell 611 and found that $\beta$ did indeed vary beyond the range of -0.2 to +0.2, and could vary significantly over the radius of a cluster. We conclude that the velocity dispersion data does not provide a meaningful constraint for our purpose of discerning core size, because a wide range of core sizes can be fit by the data with only minor variations in anisotropy, and assuming $\beta=0$ fits both large and small cores equally well. We note that the data in this case extends only to $\sim5\arcsec$, whereas the half-light radius of the BCG is $10\farcs7$, limiting the influence of the data. However, as data becomes available at larger radii and with less noise, it will offer more constraining power. Also, with more data, it may be possible to use two-dimensional kinematic analysis and/or higher order moments of the velocity dispersion to constrain the anisotropy.

\subsection{Potential Systematic Errors}
\label{Potential Systematic Errors}

Variations on the baseline models were created in order to examine the possible effects of systematic errors. One such possible source is external shear from perturbing galaxies that are close in projection to the line of sight to the lens. We found that including external shear had the effect of changing the posteriors for the halo orientation angle $\theta$ by as much as $\sim\ang{10}$, as well as the centroid coordinates $x_c$ and $y_c$, in some cases by more than 1\arcsec. However, we did not observe significant effects on the posteriors of other parameters, and including external shear did not significantly improve the fit.

The type of prior distribution can also impact the modeling results. As our baseline, we used uniform priors with wide ranges (see Table~\ref{tab:Abell_611_params1}), except for the core radius parameter, for which we used a log prior. We tested the impact of using log priors for the mass parameters of the BCG and perturbers. These did not have a significant impact on the resulting model posteriors or fit metrics.

Another source of systematic error is the triaxiality of the cluster, since lensing is only sensitive to the projected mass. Depending on the projection and axes ratios, the projected ellipticity could vary significantly. We do not yet have the models in \qlens~to take this complication into account.  This issue also becomes important when comparing to other probes such as weak lensing (sensitive to projected outer halo shape) and velocity dispersion measurements (sensitive to the 3d mass profile in the inner region)~\citep{Newman2011}. 

\subsection{Comparison to Other Works}
\label{Comparison}
Abell 611 has been studied by numerous other groups, utilizing a variety of techniques, including strong lensing, weak lensing, X-rays and stellar kinematics. Our emphasis is strong lensing, so here we focus our comparison with strong lensing results of others where possible. The work of \citetalias{Donnarumma2011}, \citet{Newman2009,Newman2013a,Newman2013b,Monna2017} are particularly relevant.

The predicted value for core size varies significantly in the literature. \citet{Monna2017} use velocity dispersion measurements of 17 cluster members in their strong lensing analysis, and infer a core size of $5.8_{-1.6}^{+2.0}$\arcsec, although they assume a dPIE profile for their halo. Their result has a reduced $\chi^2$ of 0.7 and they assume position errors of $1''$. \citetalias{Donnarumma2011} uses an NFW halo and so does not examine core size. \citet{Newman2013b} find a core size of $\sim0.7''$ in their cNFW model, in which they have a reduced $\chi^2$ of $\sim1$ and they assumed position errors of $0.5\farcs$. Our preferred solution (i.e., cNFW, small-core mode) is for a core size of <1\arcsec, with a reduced $\chi^2$ of 0.28 and assumed position errors of $0\farcs2$. 

In our preferred model (cNFW, small-core), the median posterior value of the dark matter halo mass $M_{200}$ is $\num{1.2e15}M_{\sun}$. As several of the prior analyses \citep{Donnarumma2011,Newman2009,Richard2010} used an incorrect value for the redshift of one of the sources, their strong lensing mass results are not directly comparable, so instead we compare to their weak lensing and X-ray results. The X-ray analysis of \citetalias{Donnarumma2011} found a mass of $\sim\num{1e15}M_{\sun}$. \citet{Newman2013a}, which did use the correct source redshifts, found a halo mass of $\num{8.3e14}M_{\sun}$ in their combined analysis. \citet{Romano2010} used two weak lensing techniques and various model profiles, and found $M_{200}$ to be in the range of $\num{5.3e14}M_{\sun} - \num{5.9e14}M_{\sun}$, with moderate uncertainties. In \citet{Richard2010}, their X-ray analysis puts the 2-D projected mass within $R<250$ kpc as $\num{2.06e14}M_{\sun}$, while the same statistic for our model is $\num{2.12e14}M_{\sun}$, similar to theirs. 

\subsubsection{The Importance of Position Errors}
\label{The Importance of Position Errors}

The magnitude of assumed positional errors $\sigma_{pos}$ of the observed image positions directly impacts the $\chi^2$ of the strong lensing model, as $\sigma_{pos}^2$ appears in the denominator of the equation for $\chi^2$. This in turn impacts comparisons with other modeling methods. When combining strong lensing analysis with other approaches such as weak lensing, stellar kinematics or X-ray analysis, authors often assume a strong lensing positional error that accommodates possible deficiencies in the lens models  \citep{Newman2013b,Zitrin2015}. In our case, this is not a consideration since we are only employing one type of analysis. In addition, systematic errors (see discussion in Section~\ref{Potential Systematic Errors}) are often difficult to quantify, and an attempt is sometimes made to account for those errors by increasing the assumed positional error, sometimes dramatically.

\citetalias{Donnarumma2011} assumed positional errors of $0\farcs2$, while \citet{Newman2013b} used $0\farcs5$, \citet{Monna2017} used $1\farcs0$ and \citet{Zitrin2015} used $1\farcs4$. We made a model run with the positional error as a free parameter, resulting in a best-fit value of $0\farcs18$. We ultimately adopted a position error value of $0\farcs2$. Nevertheless, the reduced $\chi^2$ of our model is quite low at 0.28 (although we did exclude images that would have raised that, as discussed in Section~\ref{Abell 611 Data}). Had we used higher values of position error such as $0\farcs5$ or $1\farcs4$, the reduced $\chi^2$ would have been 0.045 or 0.0057, respectively.

\section{Conclusions}
\label{Conclusions}

% Using novel lensing software, we perform strong lens modeling of the galaxy cluster Abell 611, as well as mock data modeling of a simulated galaxy cluster constructed to be similar in nature to Abell 611, with the goal of determining dark matter distribution in the central region. 

Our main aim in this paper was to put robust constraints on the dark matter densities in the central regions of clusters using strong lensing alone. Constraints on the central dark matter density of clusters is critical for constraining the particle physics of self-interacting dark matter models. 

We used simulated cluster data to test whether strong lensing data could distinguish between cuspy and cored data sets (with a core radius of 10\arcsec), both with and without central images present. The non-central images were in the $10"-30"$ range, in agreement with observed images. Our main findings from the analysis of mock data are as follows.

\begin{itemize}

\item It is possible to distinguish between the cored and cuspy data sets, even in absence of central images, provided the density profile and shape of the density contours are accurately modeled. For the cored halo mock data with core radius of 10\arcsec, we infer a core radius greater than $3\farcs89$ at 95\% confidence level. For the cuspy data set, we infer a core radius less than $1\farcs01$ at 95\% confidence level.

% For the cored data set, the median value of core radius (true value of 10.0\arcsec) is 7.0\arcsec, with a 1-$\sigma$ lower bound of 4.8\arcsec. For the cuspy data set (true core radius of zero), the median fit value is 0.3\arcsec, with a 1-$\sigma$ upper bound of 0.6\arcsec.

\item Approximating the potential with a pseudo-elliptical model rather than using a true elliptical density can degrade parameter recovery for strongly elliptical halos. In the case of a dark matter halo with axis ratio $q=0.5$, the halo mass and concentration parameters were both outside their 2-$\sigma$ contours. Although the inferred core size was not significantly biased in this case, combining these results with other probes such as weak lensing to better constrain the mass distribution would likely bias the inferred core size significantly, illustrating a specific danger of combining multiple probes when modeling systematics are present.

\item The use of a radial density profile with a different shape 
than that of the mock data caused the inferred surface density (hence, core size) in the regions void of images (i.e., either near the center or on the outskirts of the cluster) to be biased. We find this effect can be severe enough to make a cored halo appear cuspy and vice versa, even though the profile remains well-fit in the range of radii where the images are located. This systematic can be alleviated if visible central images are present in the data.
\end{itemize}

With these lessons, we modeled Abell 611 with two halo profiles (``cNFW" and ``Corecusp") that allow for a variable core size, a model for the BCG and seven cluster members (see Section~\ref{Abell 611 Lens Model}). Our main findings are the following. 
\begin{itemize}
\item Both the cNFW and Corecusp models found similar solutions. The cNFW model has the lower $\chi^2$ and is the preferred model with higher Bayesian evidence. Reduced $\chi^2$ values of 0.28 and 0.30 were obtained for the cNFW and Corecusp models, respectively, even with a small value of assumed position error of $0\farcs2$. 

\item A bimodal solution was found for key parameters such as core size, halo mass and BCG mass. The large-core solution did not allow for reasonable values of BCG stellar mass-to-light ratios, with $(M_*/L_V) > 10.3$ at 95\% confidence level. For the small-core solution, we found $(M_*/L_V) > 5.5$ at 95\% confidence level. The required $(M_*/L_V)$ for the large core solution is not consistent with the measurement of stellar mass-to-light ratios in ATLAS3D early-type galaxies. The required slope of the IMF for the large-core solution is also inconsistent with various inferences~\cite{conroy2017,leier2016}, as summarized in Figure~\ref{fig:ML_vs_IMF}. This evidence points to the small core as the reasonable solution for Abell 611, consistent with the finding of \citet{Newman2013b}.

\item We infer a bottom-heavy IMF for the BCG, with IMF log-slope $\Gamma \gtrsim 1.5$ (per logarithmic interval) for stellar mass below $M_{\sun}$, at 95\% C.L. Since the lensing data are most sensitive to the BCG mass within $\lnsim$ 0.2 times the half-light radius, this result is consistent with recent studies that find an extreme bottom-heavy IMF at the centers of massive elliptical galaxies.

\item Fitting the pseudo-elliptical halo model to Abell 611 results in an inferred BCG mass that is 50\% larger compared to using the true elliptical density. This inflates the stellar mass-to-light ratio significantly, despite yielding a low inferred ellipticity ($\epsilon \approx 0.19$), and illustrates the danger of fitting the pseudo-elliptical model even in cases where the inferred ellipticity may not be extreme.

% The small-core solutions also had halo concentrations more consistent with expectations from concordance $\Lambda$CDM model. 
% The ``small-core" solution is preferred, considering the resulting mass-to-light ratio of the BCG, and results in a dark matter halo core of <1\arcsec. A large-core solution with a core size of $\sim14''$ is also statistically viable, but it requires a very massive BCG (i.e., a mass-to-light ratio in V-band of $\sim13$, in conflict with current constraints on the stellar initial mass function in large elliptical galaxies.)

\item The scaled surface density ($\kappa$) at 5\arcsec is found to be $1.32\pm0.01$, and is a particularly well-constrained parameter in all models. We expect this to be a key constraint on models of self-interacting dark matter. The inferred core density and core size (for the preferred small-core solution) are consistent with those found previously by \citet{Newman2013b}, whose results were used by \citet{2016PhRvL.116d1302K} to argue that Abell 611 prefers a self-interaction cross section over mass of about $0.06^{+0.07}_{-0.03} \mathrm{cm^2/g}$ at a relative velocity of about $1500~\mathrm{km/s}$. Our robust inference of the core size in Abell 611 underscores the promise of density profile measurements in galaxy clusters to measure the self-interaction cross section of dark matter with a precision of $0.1~\mathrm{cm^2/g}$ or better.  

\item The existing kinematic data for Abell 611 only go out to about half of the half-light radius and thus are highly sensitive to the unknown velocity dispersion anisotropy parameter $\beta_{aniso}$. With more data that provide constraints on velocity dispersion to 2-3 half-light radii, we expect that the velocity dispersion constraints will play an important role in constraining the mass-to-light ratio of the BCG and hence the underlying dark matter halo profile. 
\end{itemize}

We have shown how gravitational strong lensing can be used to put robust constraints on the dark matter halo core size and core density in galaxy clusters. Our results for Abell 611 prefer a high central density and small core size. The corresponding constraint on the self-interaction cross section at velocities of about $1500~\mathrm{km/s}$ is expected to be at the $0.1~\rm cm^2/g$ level, which would be the tightest constraint on the dark matter self-interaction cross section.

\section*{Acknowledgements}

We would like to thank Drew Newman for providing valuable feedback on the manuscript and for insightful discussions during the course of the project. We gratefully acknowledge a grant of computer time from XSEDE allocation TG-AST130007. QM was supported by NSF grant AST-1615306 and MK was supported by NFW grant PHY-1620638.
	
%%%%%%%%%%%%%%%%%%%%%%%%%%%%%%%%%%%%%%%%%%%%%%%%%%



\bibliographystyle{mnras}
\bibliography{Cluster_Lensing_DM_Distributions.bib}

% \begin{thebibliography}{99}
% \bibitem[\protect\citeauthoryear{Author}{2012}]{Author2012}
% Author A.~N., 2013, Journal of Improbable Astronomy, 1, 1
% \bibitem[\protect\citeauthoryear{Others}{2013}]{Others2013}
% Others S., 2012, Journal of Interesting Stuff, 17, 198
% \bibitem{\protect\citeauthoryear{Newman}{2009}]{Newman2009}

% \end{thebibliography}

%%%%%%%%%%%%%%%%%%%%%%%%%%%%%%%%%%%%%%%%%%%%%%%%%%

%%%%%%%%%%%%%%%%% APPENDICES %%%%%%%%%%%%%%%%%%%%%

\appendix

\section{Relevant lensing formulas for cored halo models}\label{sec:appendix_a}

\subsection{Cored NFW halo model}

The cored NFW (cNFW) model is defined by modifying the NFW profile as follows:

\begin{equation}
\rho = \frac{\rho_s r_s^3}{\left(r_c + r\right)\left(r_s + r\right)^2}.
\end{equation}

Defining $x = r/r_s$ and $\beta = r_c/r_s$, by integrating the density profile along the line of sight we find an analytic expression for the projected density profile,

\begin{equation}
\kappa(x) = \frac{2\kappa_s}{(\beta-1)^2}\left\{\frac{1}{x^2-1}\left[1-\beta-(1-x^2\beta)\mathcal{F}(x)\right] - \mathcal{F}\left(\frac{x}{\beta}\right)\right\}
\label{eq:cNFW_kappa}
\end{equation},
where we have defined $\kappa_s = \rho_s r_s/\Sigma_{cr}$, and

\begin{equation}
{\mathcal F}(x) =
\begin{cases}
  \frac{1}{\sqrt{x^2-1}}\,\mbox{tan}^{-1} \sqrt{ x^2-1 } & (x>1) \\
  \frac{1}{\sqrt{1-x^2}}\,\mbox{tanh}^{-1}\sqrt{ 1-x^2 } & (x<1) \\
  1                                                      & (x=1)
\end{cases}
\end{equation}

When using the pseudo-elliptical approximation (discussed in Section \ref{Pseudo-Elliptical Approximation}), it is useful to have an analytic formula for the deflection angle generated by a spherical cNFW lens. By integrating Eq.~\ref{eq:cNFW_kappa}, we obtain

\begin{eqnarray}
\alpha(x) & = & \frac{2\kappa_s r_s}{(1-\beta)^2 x}\biggl\{(1-\beta)^2\ln\left(\frac{x^2}{4}\right) - \beta^2\ln\beta^2  ~ ~ + ~ ~ ~ ~ ~ ~ ~  \\
& & ~ ~ ~ 2(\beta^2-x^2)\mathcal{F}\left(\frac{x}{\beta}\right) + 2[1+\beta(x^2-2)]\mathcal{F}(x)\biggl\}.
\end{eqnarray}

It can be easily verified that in the limit $\beta \rightarrow 0$, these formulae reduce to the usual analytic formulas for an NFW profile \citep{Golse2002}. Numerical convergence of these formulae becomes difficult in the neighborhood of either $x \approx \beta$, $x \approx 1$ or $\beta \approx 1$; in each of these cases, series expansions can be used for greater accuracy, all of which have been implemented and tested in the \qlens~code.

\subsection{Corecusp halo model}
\label{sec:lensing_formulae}
The Corecusp model is generated by including a core in the ``cusped halo model'' from \cite{munoz2001}, such that the density profile has the form

\begin{equation}
\rho = \frac{\rho_s r_s^n}{\left(r^2+r_c^2\right)^{\gamma/2} \left(r^2+r_s^2\right)^{(n-\gamma)/2}}
\label{Corecusp_eq}
\end{equation}
where $r_c$ is the core radius and $r_s$ acts as the scale radius where the power law ``turns over''; it can also act as a tidal radius if the outer slope $n$ is chosen to be steep enough. Choosing $n=3$ corresponds to a cored Pseudo-NFW profile, while $n=4$ corresponds to the dual pseudo-isothermal ellipsoid (dPIE) profile. If we allow $\gamma$ to vary but set $n = \gamma$, the model reduces to the often-used softened power-law model (Barkana 1998). The advantage of this profile is that a scale radius (or tidal radius) is included, while still allowing for a 
variable inner slope $\gamma$ and core radius $r_c$.

If the density profile is integrated over the line-of-sight to obtain $\kappa(R|r_s,r_c)$ where $R$ is the projected radius, then it can be shown that this is equivalent to

\begin{equation}
\kappa(R|r_s,r_c) = \left(\frac{r_s}{r_s'}\right)^n\kappa_0(R'|r_s')
\label{kaps_eq}
\end{equation}
where
\begin{equation}
R' = \sqrt{R^2+r_c^2}, ~ ~ ~ r_s' = \sqrt{r_s^2-r_c^2}.
\label{subs_eq}
\end{equation}
and $\kappa_0$ is defined as the coreless model, in other words, $\kappa_0(R|r_s) \equiv \kappa(R|r_s,r_c=0)$. Thus, the cored kappa profile can be obtained from the coreless ($r_c=0$) profile, for which the kappa and deflection formulas are known and given in Munoz et al. 2001, using the above substitutions.  To simplify the notation, we will simply write $\kappa(R|r_s,r_c)$ as $\kappa(R)$, and define

\begin{equation}
\tilde\kappa(R) \equiv \kappa_0(R|r_s'=\sqrt{r_s^2-r_c^2}),
\end{equation}
so using this notation we rewrite eq.~\ref{kaps_eq} as

\begin{equation}
\kappa(R) = \left(1-\frac{r_c^2}{r_s^2}\right)^{-\frac{n}{2}}\tilde\kappa(\sqrt{R^2+r_c^2})
\end{equation}

The corresponding radial deflections will be referred to as $\tilde\alpha(R')$ and $\alpha(R)$, again using the same variable substitutions; in other words, we define

\begin{equation}
\tilde\alpha(R) \equiv \alpha_0(R|r_s'=\sqrt{r_s^2-r_c^2}).
\end{equation}

It is important to keep in mind that whenever we evaluate $\tilde\kappa$ and $\tilde\alpha$, we must make the transformation $r_s \rightarrow \sqrt{r_s^2-r_c^2}$ in the formulas for the corresponding coreless model. 

To obtain the formula for the radial deflection, we use

\begin{eqnarray}
\alpha(R) & = & \frac{2}{R}\int_0^R u\kappa(u)du \nonumber \\
& = & \left(\frac{r_s}{r_s'}\right)^n \frac{2}{R}\int_0^R u \tilde\kappa(\sqrt{u^2+r_c^2})du ~ ~ ~ \textrm{(using eqs.~\ref{kaps_eq}, \ref{subs_eq})} \nonumber \\
& = & \left(\frac{r_s}{r_s'}\right)^n \frac{2}{R}\int_{r_c}^{\sqrt{R^2+r_c^2}} 
w \tilde\kappa(w)dw
\end{eqnarray}
and hence,
\begin{equation}
\alpha(R) = \left(1-\frac{r_c^2}{r_s^2}\right)^{-\frac{n}{2}} \left[ \frac{\sqrt{R^2+r_c^2}}{R}\tilde\alpha(\sqrt{R^2+r_c^2}) - \frac{r_c}{R} \tilde\alpha(r_c) \right].
\end{equation}

Thus we find that the radial deflection of the cored profile can be expressed as a linear combination of radial deflections from the corresponding coreless profile, again using the substitutions in eq.~\ref{subs_eq}. Thus, the same formulas for $\kappa$ and $\alpha$ from Munoz et al. (2001) can be employed for the cored model using the above transformations.

The above transformations can be easily verified for the dPIE case ($n=4$, $\gamma=2$), where $r_s$ is interpreted as a tidal radius, yielding:

\begin{equation}
\kappa = \frac{b}{2} \left[ \frac{1}{(r_c^2 + R^2)^{1/2}} - \frac{1}{(r_s^2 + R^2)^{1/2}} \right],
\end{equation}
\begin{equation}
\alpha = \frac{b}{R}\left[\sqrt{r_c^2+R^2} - r_c - \sqrt{r_s^2+R^2} + r_s\right]
\end{equation}
where
\begin{equation}
b \equiv \frac{2\pi\rho_s}{\Sigma_{cr}}\frac{r_s^4}{r_s^2-r_c^2}.
\end{equation}

As an quick check, note that if the core radius $r_c$ is set to zero, we can apply the transformations in eqs.~\ref{kaps_eq} and \ref{subs_eq} to recover the same formulas for the cored profile and its corresponding deflection angle.

\onecolumn
\FloatBarrier
\section{Table of Model Parameters}
\label{param_tables}

\begin{table}
	\centering
    \caption{The values of the fixed parameters for the BCG and seven perturbers in the Abell 611 Lens Model. The magnitude values are in the ST magnitude system. The magnitude of the object is used to determine its core radius, cutoff radius and mass parameters via the scaling relations described in the text.}
		\begin{tabular}{ccccccc} % columns, alignment for each
		\hline
		Cluster Member No.  & Core Radius & Axis Ratio & Orientation & x & y &  $m_{606w}$ \\
                      & (arc sec) & & (degrees) & (arc sec) & (arc sec) & (mag)\\
           &  \\              
		\hline
        BCG &  0.0555    & 0.70 & 132.5 & 0.0    &  0.0 & 17.0 \\
		1      & 0.0101 & 0.83 & 112.8 & 2.33   & -7.85 & 20.7 \\
        2      & 0.0067 & 0.92 & 13.1  & 3.14   & -10.05 & 21.6 \\
        3      & 0.0101 & 0.50 & 78.7  & -5.15  & 17.42 & 20.7 \\
        4      & 0.0096  & 0.67 & 80.6  & -10.88 & 10.22 & 20.8 \\
        5      & 0.0073 & 0.84 & 128.4 & -16.79 & 0.60 & 21.4 \\
        6      & 0.0055 & 0.90 & 131.6 & 1.13   & -2.78 & 22.0 \\
        7      & 0.0084 & 0.79 & 61.7  & -13.68 & 12.87 & 21.1 \\
        ref. galaxy & 0.0350 & -   & -   & -   & -     & 18.0 \\
		\hline
	\end{tabular}
	\label{fig:Abell_611_params2}
    
\end{table}
\FloatBarrier

\section{Triangle Plots of Posterior Distributions}
\label{triangle_plots}
\begin{figure*}
    \caption{Posterior distributions and two-dimensional correlations using mock data for the cored case without central images. True parameter values are indicated in orange.}
    \includegraphics[width=\linewidth]{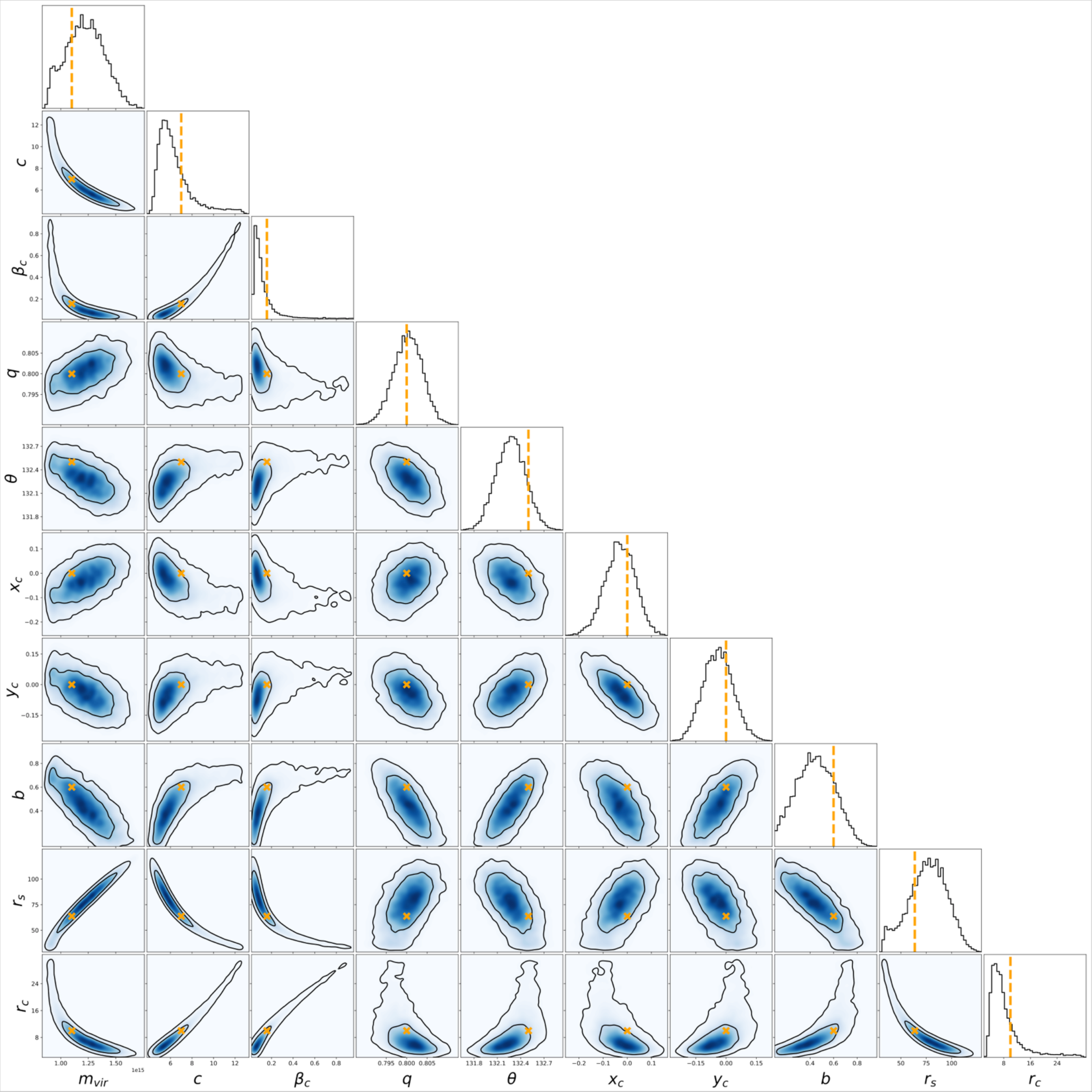}
    \label{fig:cored_triangle}
\end{figure*}

\begin{figure*}
    \caption{Posterior distributions and two-dimensional correlations using mock data for the cuspy case without central images. True parameter values are indicated in orange.}
    \includegraphics[width=\linewidth]{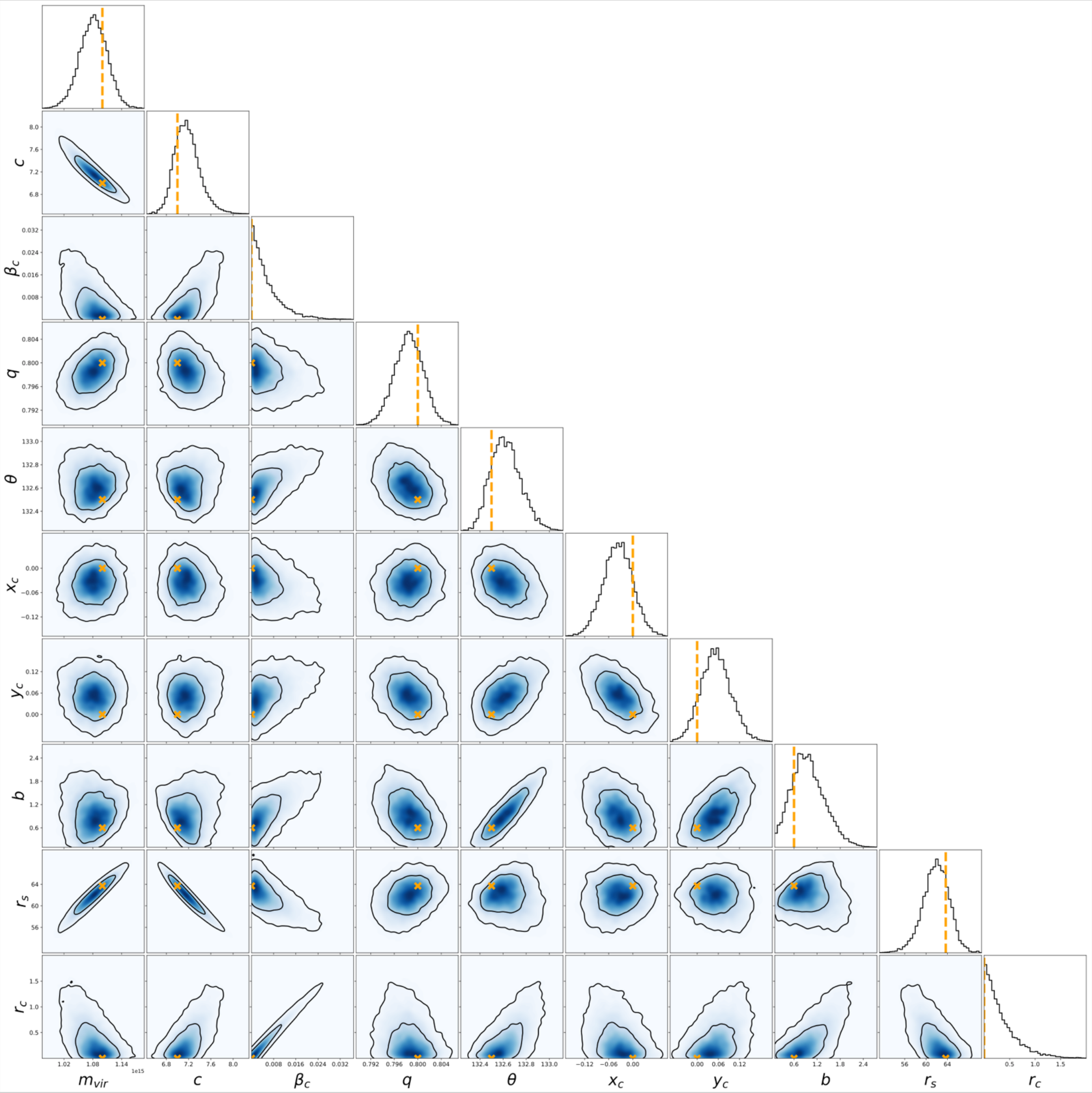}
    \label{fig:cuspy_triangle}
\end{figure*}

\begin{figure*}
    \caption{Selected Abell 611 posterior distributions and two-dimensional correlations for the cored NFW case.}
`'    \includegraphics[width=\linewidth]{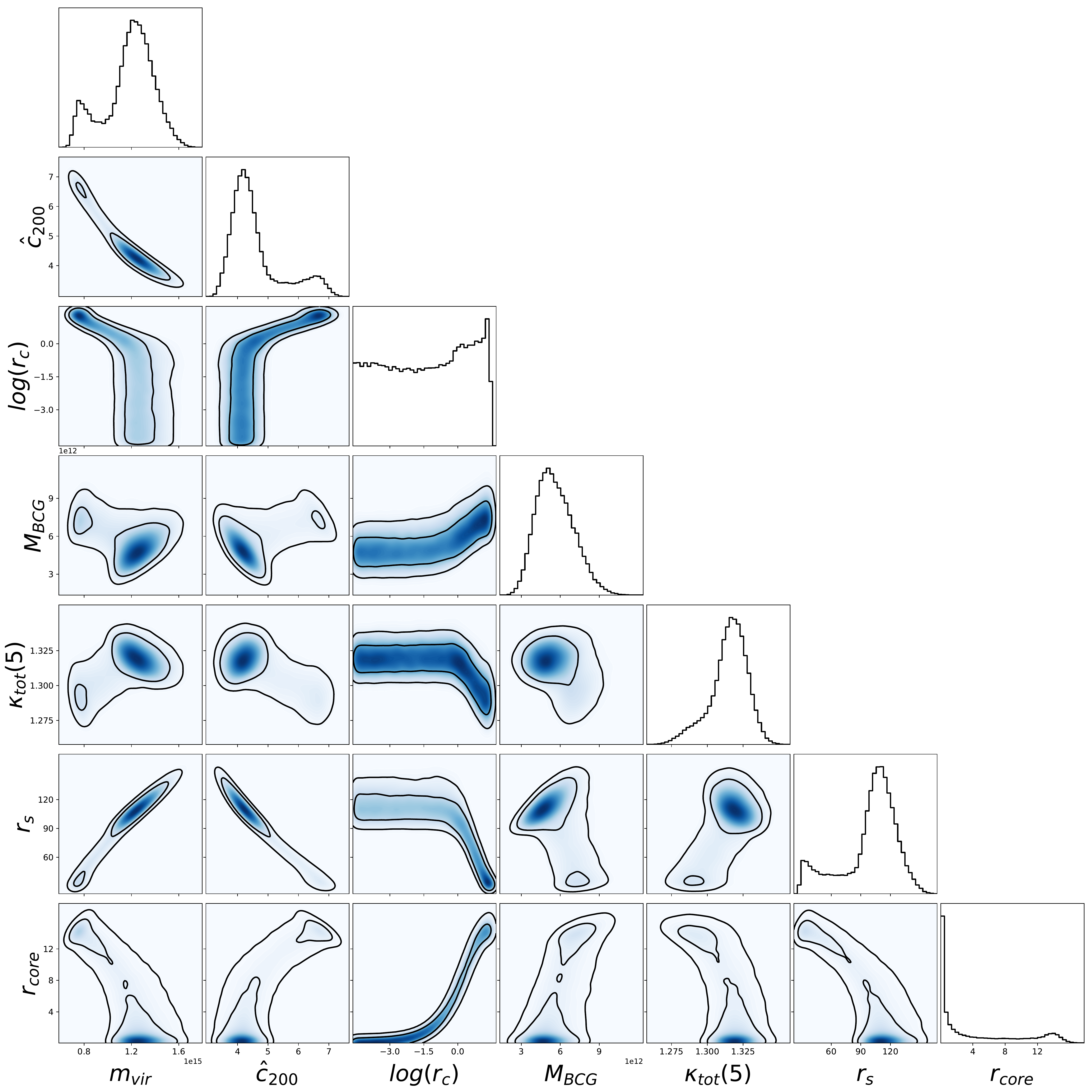}
    \label{fig:A611_cNFW_triangle}
\end{figure*}

\begin{figure*}
    \caption{Selected Abell 611 posterior distributions and two-dimensional correlations for the Corecusp case.}
    \includegraphics[width=\linewidth]{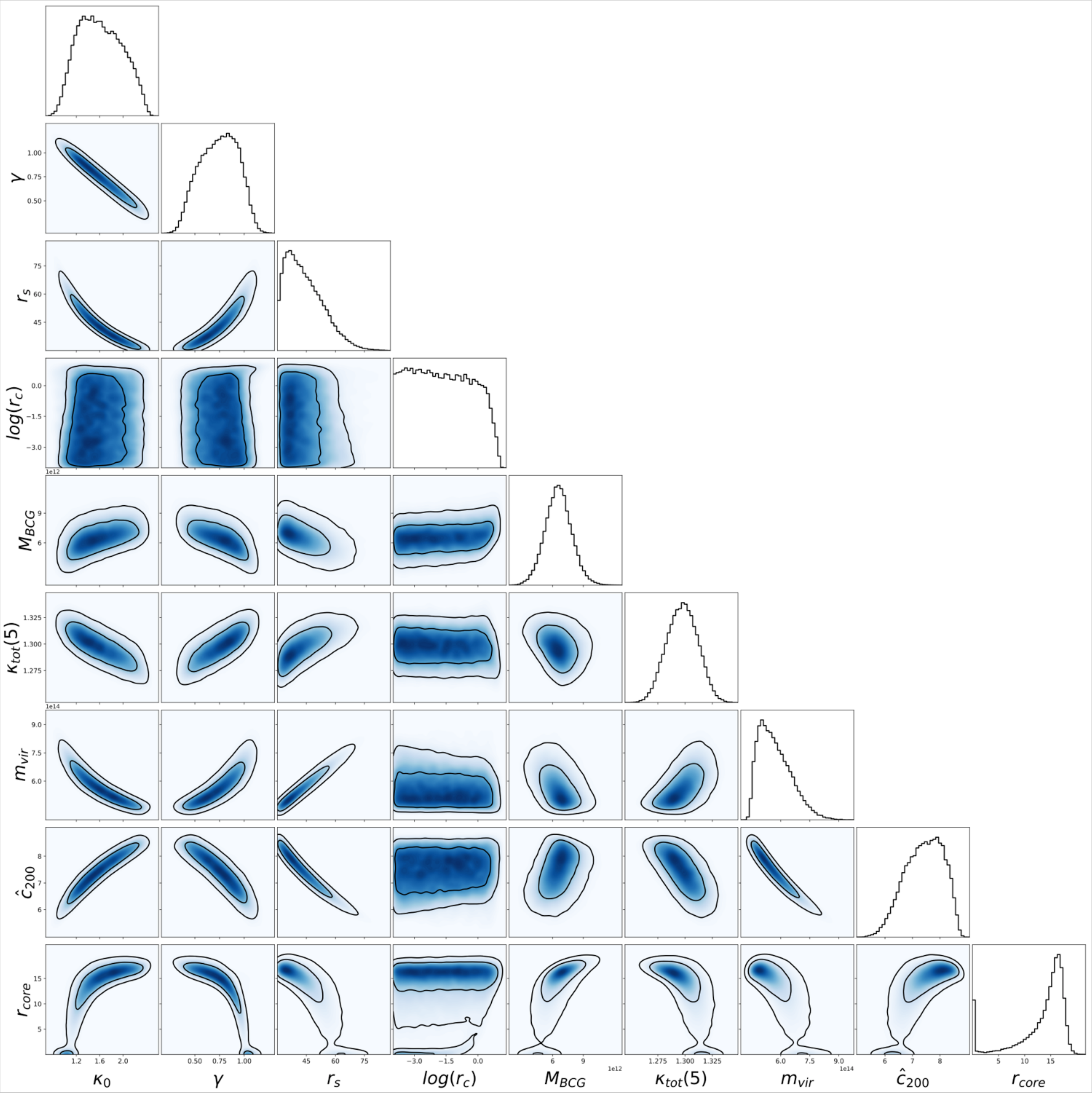}
    \label{fig:A611_cc_triangle}
\end{figure*}

%%%%%%%%%%%%%%%%%%%%%%%%%%%%%%%%%%%%%%%%%%%%%%%%%%

% Don't change these lines
\bsp	% typesetting comment
\label{lastpage}
\end{document}